\documentclass[]{pasj01}
\draft

\usepackage{color}
\usepackage[normalem]{ulem}

\usepackage{multirow}

\begin{document} 
\Received{}%{yyyy/mm/dd}
\Accepted{}%{yyyy/mm/dd}
%\Published{yyyy/mm/dd}

\title{Molecular and Atomic Clouds toward the Wolf-Rayet Nebula NGC~2359: Possible Evidence for Isolated High-Mass Star Formation Triggered by a Cloud-Cloud Collision}

%%% begin:list of authors
% Do NOT capitalize all letters in "textsc".
\author{Hidetoshi \textsc{SANO}\altaffilmark{1,2}}%
\author{Katsuhiro \textsc{HAYASHI}\altaffilmark{2}}%
\author{Rei \textsc{ENOKIYA}\altaffilmark{2}}%
\author{Kazufumi \textsc{TORII}\altaffilmark{3}}%
\author{Shun \textsc{SAEKI}\altaffilmark{2}}%
\author{Kazuki \textsc{OKAWA}\altaffilmark{2}}%
\author{Kisetsu \textsc{TSUGE}\altaffilmark{2}}%
\author{Daichi \textsc{TSUTSUMI}\altaffilmark{2}}%
\author{Mikito \textsc{KOHNO}\altaffilmark{2}}%
\author{Yusuke \textsc{HATTORI}\altaffilmark{2}}%
\author{Shinji \textsc{FUJITA}\altaffilmark{2}}%
\author{Satoshi \textsc{YOSHIIKE}\altaffilmark{2}}%
\author{Ryuji \textsc{OKAMOTO}\altaffilmark{2}}%
\author{Atsushi \textsc{NISHIMURA}\altaffilmark{4}}%
\author{Akio \textsc{OHAMA}\altaffilmark{2}}%
\author{Takahiro \textsc{HAYAKAWA}\altaffilmark{2}}%
\author{Hiroaki \textsc{YAMAMOTO}\altaffilmark{2}}%
\author{Kengo \textsc{TACHIHARA}\altaffilmark{2}}%
\author{Cristina Elisabet \textsc{CAPPA}\altaffilmark{5,6}}%
\author{Yasuo \textsc{FUKUI}\altaffilmark{1,2}}%
\email{sano@a.phys.nagoya-u.ac.jp}
\altaffiltext{1}{Institute for Advanced Research, Nagoya University, Furo-cho, Chikusa-ku, Nagoya 464-8601, Japan}
\altaffiltext{2}{Department of Physics, Nagoya University, Furo-cho, Chikusa-ku, Nagoya 464-8601, Japan}
\altaffiltext{3}{Nobeyama Radio Observatory, Minamimaki-mura, Minamisaku-gun, Nagano 384-1305, Japan}
\altaffiltext{4}{Department of Physical Science, Graduate School of Science, Osaka Prefecture University, 1-1 Gakuen-cho, Naka-ku, Sakai, Osaka 599-8531, Japan}
\altaffiltext{5}{Instituto Argentino de Radioastronom\'ia, CONICET, Argentina.}
\altaffiltext{6}{Facultad de Ciencias Astron\'omicas y Geof\'isicas, Universidad Nacional de la Plata, Paseo del Bosque s/n, 1900, La Plata, Argentina.}


\maketitle

\begin{abstract}
NGC~2359 is an H{\sc ii} region located in the outer Galaxy that contains the isolated Wolf-Rayet (WR) star HD~56925. We present millimeter/submillimeter observations of $^{12}$CO($J$ = 1--0, 3--2) line emission toward the entire nebula. We identified that there are three molecular clouds at {$V_\mathrm{LSR}$} $\sim$37, $\sim$54, {and $\sim$67 km s$^{-1}$}, and {three} H{\sc i} clouds{: two of them are at $V_\mathrm{LSR}$ $\sim$54 km s$^{-1}$ and the other is at $\sim$63 km s$^{-1}$. These clouds except for the CO cloud at 67 km s$^{-1}$ are limb-brightened in the radio continuum, suggesting part of each cloud has been ionized. We newly found an expanding gas motion of CO/H{\sc i}, whose center and expansion velocities are $\sim$51 and $\sim$4.5 km s$^{-1}$, respectively. This is consistent with large line widths of the CO and H{\sc i} clouds at 54 km s$^{-1}$. The kinematic temperature of CO clouds at 37 and 54 km s$^{-1}$ are derived to be 17 and 61 K, respectively, whereas that of the CO cloud at 67 km s$^{-1}$ is only 6 K, indicating that the former two clouds have been heated by strong UV radiation. We concluded that the 37 and 54 km s$^{-1}$ CO clouds and three H{\sc i} clouds are associated with NGC~2359, even if these clouds have different velocities. Although the velocity difference including the expanding motion} are {typical signatures of the stellar feedback from the exciting star, o}ur analysis {revealed that the observed large momentum for the 37 km s$^{-1}$ CO cloud} cannot be explained {only} by {the total wind momentum of the WR star and its progenitor. We therefore propose {an alternative} scenario that the} isolated high-mass {progenitor of HD~56925 was formed} by a collision between {the CO clouds at 37 and 54 km s$^{-1}$. If we apply the collision scenario,} NGC~2359 {corresponds to the final phase} of the cloud-cloud collision.
\end{abstract}

\section{Introduction} \label{sec:intro}
It is a longstanding question how isolated (or so-called ``field'') high-mass stars are formed in interstellar space{, because high-mass stars are thought to be formed as a cluster (e.g., \cite{2018ARA&A..56...41M} and references therein).} Recently, the cloud-cloud collision model has received considerable attention as a formation mechanism not only for high-mass stellar clusters but also for isolated high-mass stars. {\citet{2015ApJ...806....7T}} showed that the isolated O-type star associated with the H{\sc ii} region {RCW~120} was formed by a collision between two molecular clouds, that have a velocity separation of {$\sim$20 km s$^{-1}$}. {The high intensity ratios ($> 0.8$) of CO $J$ = 3--2 / 1--0 toward both clouds correspond to a kinetic temperature of $\gtrsim 20$--40 K, indicating radiative heating of the two clouds by the O-type star.} They also found that the two clouds have complementary spatial distributions and that a bridging feature {of H{\sc i}} physically connects the two {CO} clouds in velocity space. {The authors} concluded that the complementary spatial distributions and the bridging feature can be interpreted as typical cloud-cloud collision signatures. Therefore, the isolated O-type star in {RCW~120} was formed by strong gas compression during the cloud-cloud collision.

Theoretical studies also support this {idea of triggered star formation} because a collision between two dense clouds increases the effective sound speed and gas density in the shocked layer (e.g., \cite{1992PASJ...44..203H,2010MNRAS.405.1431A,2013ApJ...774L..31I,2014ApJ...792...63T,2018PASJ...70S..53I}). According to \citet{2018PASJ...70S..53I}, isothermal magnetohydrodynamic simulations of cloud-cloud collisions can achieve high {mass} accretion rates $M_\mathrm{acc}$ $>$ $10^{-4}$ $M_\odot$ yr$^{-1}$, which is high enough to allow the formation of O-type stars (e.g., \cite{1987ApJ...319..850W}). They also found that the most massive sink particle, which has a mass of 50 $M_\odot$ or higher, was created within a few $10^5$ yr after the collision.

{As of early 2020}, observational evidence for cloud-cloud collisions as a formation mechanism for O-type {(or early B-type)} stars has been obtained from {34} sources associated with stellar clusters containing {more than two high-mass} stars {(DR~21, \cite{2019PASJ...71S..12D}; Sgr~B2, \cite{1994ApJ...429L..77H}; W~49~N, \cite{1993ApJ...413..571S}; Arches cluster, \cite{2008ApJ...675.1278S}; Westerlund~2, \cite{2009ApJ...696L.115F}, \cite{2010ApJ...709..975O}; NGC~3603, \cite{2014ApJ...780...36F}; 50~km s$^{-1}$ cloud in Sgr A, \cite{2015PASJ...67..109T}; RCW~38, \cite{2016ApJ...820...26F}; R~136, \cite{2017PASJ...69L...5F}; Sh2-237, \cite{2017ApJ...834...22D}; N37 \& G35.20$-$0.74, \cite{2017ApJ...837...44D}; E-S235ABC, \cite{2017ApJ...849...65D}; N49, \cite{2017ApJ...851..140D}; M42, \cite{2018ApJ...859..166F}; NGC~6334 \& NGC~6357, \cite{2018PASJ...70S..41F}; NGC~6618, \cite{2018PASJ...70S..42N}; RCW~36, \cite{2018PASJ...70S..43S}; RCW~79, \cite{2018PASJ...70S..45O}; S116, S117 \& S118, \cite{2018PASJ...70S..46F}; N21 \& N22, \cite{2018PASJ...70S..47O}; W~33, \cite{2018PASJ...70S..50K}; NGC~604, \cite{2018PASJ...70S..52T}; G24.80$+$0.10, \cite{2018ApJ...866...20D}; LHA~120-N~44, \cite{2019ApJ...871...44T}; AFGL~5142, \cite{2019ApJ...875..138D}; G49.5$-$0.4 \& G49.4$-$0.3, \cite{2019PASJ..tmp...46F}; M33GMC~37, \cite{2019arXiv190808404S}; Firecracker in Antennae Galaxies, \cite{2019arXiv190905240T})}, and with {27} sources associated with an isolated high-mass star {(IRAS~04000$+$5052, \cite{2004ApJ...614L.105W}; BD~$+$40~4124, \cite{2006ApJ...642..330L}; S87, \cite{2008ApJ...680..446X}; M20, \cite{2011ApJ...738...46T}, \yearcite{2017ApJ...835..142T}; RCW~120, \cite{2015ApJ...806....7T}; N~159~W-South, \cite{2015ApJ...807L...4F}, \cite{2019ApJ...886...15T}; N~159~E-Papillon, \cite{2017ApJ...835..108S}, \cite{2019ApJ...886...14F}; E-S235 main East1 \& Southwest, \cite{2017ApJ...849...65D}; M43, \cite{2018ApJ...859..166F}; GM~24, \cite{2018PASJ...70S..44F}; G018.149$-$0.283, \cite{2018PASJ...70S..47O}; RCW~34, \cite{2018PASJ...70S..48H}; RCW~32, \cite{2018PASJ...70S..49E}; N35, G024.392$+$00.072 \& G024.510$-$00.060, \cite{2018PASJ...70S..51T}; Sh2-48, \cite{2018PASJ..tmp..121T}; S44, \cite{2018PASJ..tmp..126K}; N4, \cite{2019ApJ...872...49F}; G49.57$-$0.27, \cite{2019PASJ..tmp...46F}; IRAS~18223$-$1243, \cite{2018ApJ...861...19D}; G24.85$+$0.09, G24.74$+$0.08, G24.71$-$0.13, and G24.68$-$0.16, \cite{2018ApJ...866...20D}; G8.14$+$0.23, \cite{2019ApJ...878...26D}).}

However, isolated Wolf-Rayet (WR) stars have not previously been studied as products of star formation triggered by {a} cloud-cloud collision. A WR star is thought to be a late evolutionary stage of an O-type star with a mass of $\sim$25 $M_\odot$ or higher. WR stars have lost their hydrogen envelopes via strong stellar winds, with typical velocities and mass-loss rates of $\sim$1,000 km s$^{-1}$ to 3,000 km s$^{-1}$ and $\sim$10$^{-5}$ $M_\odot$ yr$^{-1}$, respectively (e.g., \cite{2007ARA&A..45..177C} and references therein). {To explore the origin of an isolated high-mass star that can become a WR star,} we need to clarify whether the formation of {the progenitor of} isolated WR star may have been {also} triggered by a cloud-cloud collisions.

\begin{figure}
\begin{center}
\includegraphics[width=90mm]{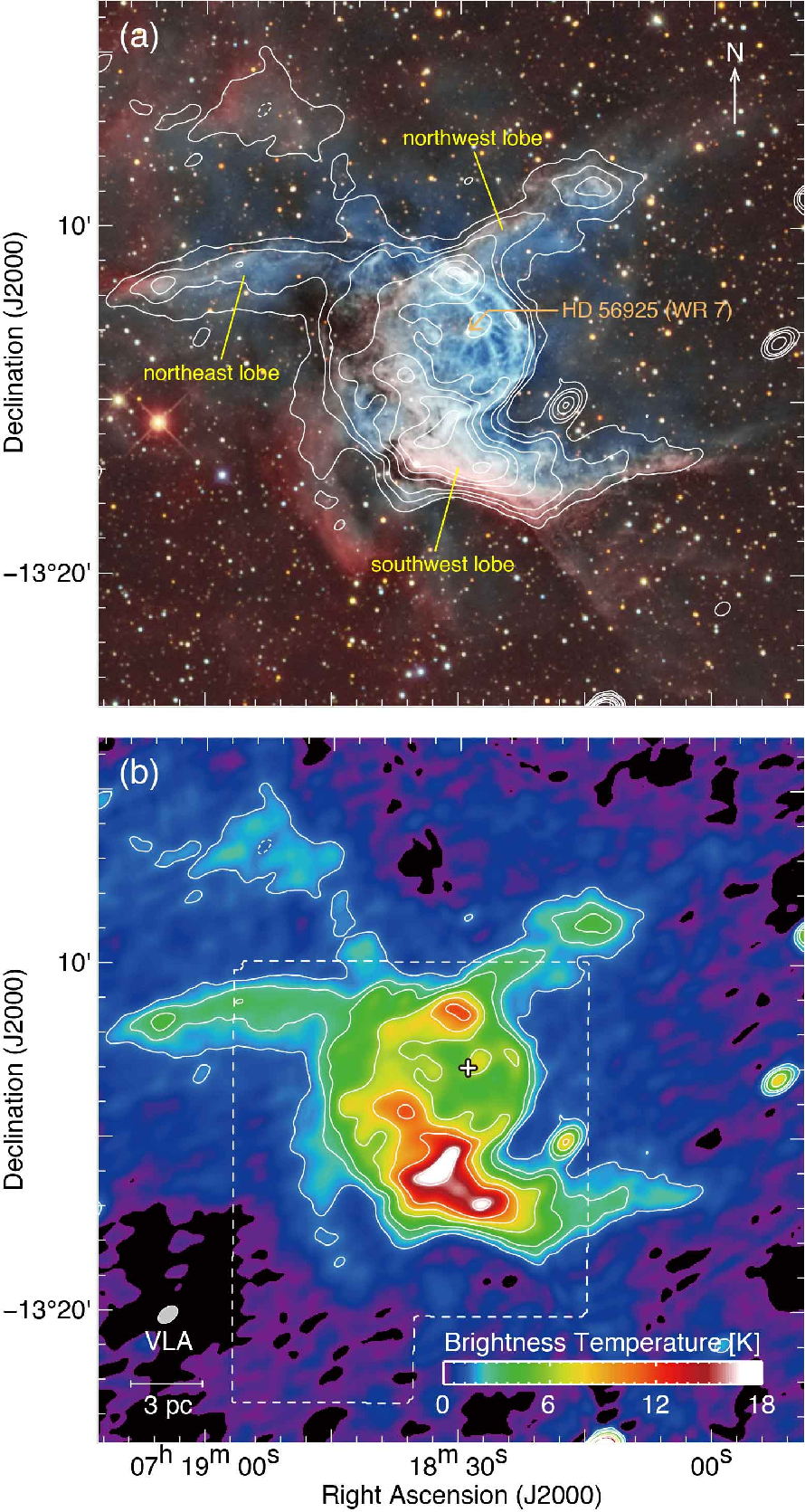}
\end{center}
\caption{(a) Optical image of the Wolf-Rayet nebula NGC~2359 (courtesy of Robert Franke) obtained at the Focal Pointe Observatory. The red and green/blue represent H$\alpha$ and {[O{\sc iii}] line emission}, respectively. Superposed contours indicate the radio continuum with VLA \citep{1999AJ....118..948C}. The contour levels are 1.5, {2.5, 3.5, 6.0, 8.5, 11.0, and 16.0} K. {(b) Radio continuum image superposed on its intensity contours. The contour levels are the same as in Figure \ref{top}(a). The beam size of radio continuum and scale bar at the distance of 5 kpc are also shown in the left bottom corner. The dashed polygon indicates the observed area of $^{12}$CO($J$ = 3--2) line emission. The position of the WR star is also indicated by arrow or cross symbol.}}
\label{top}
\end{figure}

NGC~2359 (also known as Sh 2--298 or Thor's Helmet) is an optical ring nebula associated with the isolated WR star HD~56925 ($=$ WR~7, \cite{1981SSRv...28..227V}). Figure \ref{top}{(a)} shows an optical image of NGC~2359 (courtesy of Robert Franke) obtained at the Focal Pointe Observatory. The WR star is categorized as spectral type WN4b (\cite{1996MNRAS.281..163S}) and is located at {($\alpha_\mathrm{J2000}$, $\delta_\mathrm{J2000}$) = ($07^{\mathrm{h}}18^{\mathrm{m}}29\fs13$, $-13{^\circ}13\arcmin01\farcs5$)}. It has formed a wind-blown bubble {($\sim$4\farcm5 in diameter) with bright optical lobes in the directions of northwest, southwest, and northeast (as shown in Figure \ref{top}{a)}}. By using the interferometric profiles of H$\alpha$, [O{\sc iii}], and [N{\sc ii}], the expansion velocity of the bubble was found to range from $\sim$15 to 30 km s$^{-1}$ \citep{1981ApJ...243..184S,1982ApJ...254..569T,1983A&A...117..127G,1994A&A...285..631G,1994Ap&SS.216..281M} up to 55 $\pm$ 25 km s$^{-1}$ \citep{1973SvA....17..317L}. Two prominent optical dark lanes are located at {($\alpha_\mathrm{J2000}$, $\delta_\mathrm{J2000}$) $\sim$ ($07^{\mathrm{h}}18^{\mathrm{m}}36\fs84$, $-13{^\circ}16\arcmin46\arcsec$) and ($07^{\mathrm{h}}18^{\mathrm{m}}44\fs15$, $-13{^\circ}12\arcmin26\arcsec$). The nebula is also bright in radio continuum. Figure \ref{top}(b) shows the radio continuum image with the Very Large Array (VLA; \cite{1999AJ....118..948C}). The radio continuum distribution is roughly consistent with the optical image including the bright three lobes and the wind-blown bubble.}

The distance to the nebula and the WR star is still a matter of debate. The photometric distance to HD~56925 was determined to be $\sim$3.5 kpc to 6.9 kpc \citep{1968MNRAS.141..317S,1971MNRAS.153..303C,2001NewAR..45..135V}, whereas the kinematic distance to NGC~2359 was found to be 4.0--6.3 kpc according to the radial velocity of the CO and optical {line emission} \citep{1973A&A....25..337G,1978ApJ...220..516P,1979ApJ...233..888T,1984ApJ...279..125F}. In the present study, we adopt an average distance of $\sim$5 kpc, following the same assumptions as in previous studies (e.g., \cite{2001AJ....121.2664C,2001A&A...366..146R}).

Both CO and H{\sc i} observations have been reported in the direction of NGC~2359. \citet{1981ApJ...243..184S} first observed the $^{12}$CO and $^{13}$CO($J$ = 1--0) {line emission} by using the NRAO 11 m radio telescope, which has an angular resolution $\Delta \theta$ $\sim$1\farcm1. {The authors} found that there are three velocity components at $V_\mathrm{LSR}$ $\sim$37, $\sim$54, and $\sim$67 km s$^{-1}$. \citet{1999AJ....118..948C} made a complete observation of H{\sc i} by using VLA at $\Delta \theta$ $\sim$45$''$. {\authorcite{1999AJ....118..948C}} found that the {H{\sc i}} components at $V_\mathrm{LSR}$ $\sim$54 and $\sim$63 km s$^{-1}$ clearly trace along the optical wind-blown bubble, and they concluded that these H{\sc i} clouds are associated with NGC~2359. Subsequent follow-up CO($J$ = 1--0, 2--1) observations were conducted using the {Swedish-ESO Submillimetre Telescope (SEST)} with a {modest} angular resolution of 22$''$--44$''$ \citep{2001AJ....121.2664C}. On the basis of a comparative study of the CO, H{\sc i}, and optical line emission, {the authors} conclude that the CO and H{\sc i} components at $V_\mathrm{LSR}$ $\sim$54 km s$^{-1}$ are definitely associated with the nebula. \citet{2001A&A...366..146R} provided further support for this conclusion. By observing the fully-sampled CO($J$ = 1--0, 2--1) {line} emission via the NRAO 12 m radio telescope ($\Delta \theta$ = 27$''$--54$''$), they found that the CO cloud shows line broadening $>$ 5.5 km s$^{-1}$ and a high kinetic temperature of up to 80 K. Furthermore, \citet{2003A&A...411..465R} determined the detailed physical conditions in the CO cloud at $V_\mathrm{LSR}$ $\sim$54 km s$^{-1}$ by using the IRAM 30 m radio telescope ($\Delta \theta$ $\sim$12$''$) with fully sampled CO($J$ = 1--0, 2--1) {line emission}. {\authorcite{2003A&A...411..465R}} found multi-shocked layers in the CO cloud at $V_\mathrm{LSR}$ = 42--48, 48--52, and 52--57 km s$^{-1}$, which may have been formed by several energetic events. {The interpretation is consistent with detections of the H$_2$ 1--0 S(1) emission (\cite{1995Ap&SS.224..271S}, \yearcite{1998AJ....115.2475S}) and the infrared shell (e.g., \cite{1996AJ....112.2828M}).} However, {there is no submillimeter observation such as CO($J$ = 3--2) line emission which can easily trace the warm molecular clouds heated by strong UV radiation. In addition to this, the physical properties of molecular clouds---kinematic temperature and number density of molecular hydrogen---have not been revealed except for the CO cloud at the velocity of $\sim$54 km s$^{-1}$.} Moreover, no study has determined a relation between the molecular clouds and formation {mechanism} of the {isolated O-type progenitor} of the WR star.

In this study, we {carried out millimeter/submillimeter observations of} CO($J$ = 1--0, 3--2) {line emission} {toward the entire molecular clouds and compared them with the archived H{\sc i} and radio continuum datasets, in order} to investigate {the physical properties of molecular clouds and their relation to the formation mechanism of the WR star.} Section \ref{sec:obs} describes the CO observations and other wavelength datasets. Section \ref{sec:results} comprises {two} subsections. Subsections \ref{sec:ngc2359} describes the distributions of molecular and atomic clouds in the direction of NGC~2359; Subsection \ref{sec:phys} presents the physical condition in the molecular clouds. The discussion and our conclusions are given in Sections \ref{sec:discuss} and \ref{sec:conc}, respectively.

\section{Observations {and Datasets}} \label{sec:obs}
\subsection{CO}
We conducted observations of $^{12}$CO($J$ = 3--2) {line emission} during November and December 2015 by using {the Atacama Submillimeter Telescope Experiment (ASTE; \cite{2004SPIE.5489..763E})}. We used the Nyquist-sampled on-the-fly (OTF) mapping mode \citep{2008PASJ...60..445S}. The observation area was $\sim$120 arcmin$^2${, which is shown in Figure \ref{top}(b)}. The front end was the cartridge-type 2SB mixer receiver ``DASH 345.'' The typical system temperature was $\sim$250 K in a single sideband. The back end system ``MAC'' used for spectroscopy \citep{2000SPIE.4015...86S}, had 1,024 channels with a bandwidth of 128 MHz. The velocity resolution and coverage were $\sim$0.11 km s$^{-1}$ {per ch} and $\sim$111 km s$^{-1}$, respectively. We convolved the data cube with a Gaussian kernel, and the final beam size was $\sim$25$''$. The pointing accuracy was checked every hour to achieve an offset within 3$''$. We calibrated the absolute intensity by observing IRC$+$10216 [$\alpha_\mathrm{B1950}$ = $9^{\mathrm{h}}45^{\mathrm{m}}14\fs8$, $\delta_\mathrm{B1950}$ = $-13{^\circ}30\arcmin40\arcsec$] \citep{1994ApJS...95..503W}, and the estimated error was less than 7$\%$. We obtained a data cube with a noise fluctuation of $\sim$0.1 K at a velocity resolution of 0.4 km s$^{-1}$.

We {also} performed $^{12}$CO($J$ = 1--0) observations from March 2015 to April 2015 by using the Nobeyama 45 m radio telescope, which is operated by Nobeyama Radio Observatory in Japan{, a branch of the National Astronomical Observatory of Japan (NAOJ).} We observed an area of $\sim$170 arcmin$^2$ by using the Nyquist sampled OTF mapping mode. The front end was the two-beam, waveguide-type, dual-polarization, sideband-separating (2SB) SIS receiver system ``TZ1'' \citep{2013PASP..125..252N}. The backend system, the Spectral Analysis Machine for the 45 m telescope (SAM45; \cite{2011GASS..1,2012PASJ...64...29K}), had 4,096 channels with a bandwidth of 250 MHz, corresponding to a velocity coverage of $\sim$650 km s$^{-1}$ and a velocity resolution of $\sim$0.18 km s$^{-1}$ {per ch}. Typical system temperatures were $\sim$300 K to 500 K for the H polarization and $\sim$400 K to 600 K for the V polarization, including the atmosphere. The data cube was smoothened with a Gaussian kernel and the final beam size was 25$''$. We checked the pointing accuracy every hour to achieve an offset less than 2$''$. We calibrated the absolute intensity by observing Orion-IRC2 [$\alpha_\mathrm{B1950}$ = $5^{\mathrm{h}}32^{\mathrm{m}}47\fs0$, $\delta_\mathrm{B1950}$ = $-5{^\circ}24\arcmin23\farcs0$], and the estimated error was $\sim$7$\%$. {Finally, we combined the cube data with the archived $^{12}$CO($J$ = 1--0) data obtained as part of the FUGIN project (FOREST Unbiased Galactic plane Imaging survey with the Nobeyama 45-m telescope, \cite{2017PASJ...69...78U}) using the root mean square weighting method. The final noise fluctuation was $\sim$0.34 K at a velocity resolution of 0.65 km s$^{-1}$.}

{Additionally, the archived $^{13}$CO($J$ = 1--0) dataset obtained with the Nobeyama 45-m ratio telescope was used for estimating the physical properties of molecular clouds. The observations were also performed as part of the FUGIN project. The initial angular resolution of the datasets is 20$\farcs7$. To improve the signal to noise ratio, we smoothed the dataset with a Gaussian kernel, and the final beam size was $\sim$25$''$. The final noise fluctuation in the data was $\sim$0.37 K at a velocity resolution of 0.65 km s$^{-1}$. } 

We also used the $^{12}$CO($J$ = 1--0) data obtained with the NANTEN 4 m millimeter/submillimeter radio telescope of Nagoya University at Las Campanas Observatory in Chile. The observations were conducted as part of the NANTEN Galactic Plane CO Survey \citep{2004ASPC..317...59M}. The telescope had a beam size of 2\farcm6 at a frequency of 115 GHz. The velocity resolution and coverage were 0.5 and 600 km s$^{-1}$, respectively. In this paper, the beam size {was smoothed} to 4$'$. The final noise fluctuation {of} the data was $\sim$0.24 K at a velocity resolution of 0.5 km s$^{-1}$.

\subsection{H{\sc i} {and} Radio Continuum}
To {reveal} the spatial distributions of atomic hydrogen gas and ionized matter, we used the datasets for H{\sc i} and for the radio continuum at 1465 MHz that appear in \citet{1999AJ....118..948C}. The data were obtained using the VLA of the National Radio Astronomy Observatory in April and August 1996. The final beam size {were} 57\farcs7 $\times$ 40\farcs6 with a position angle of $-$16$^{\circ}$ for the H{\sc i}; 39\farcs2 $\times$ 24\farcs9 with a position angle of $-$52$^{\circ}$ for the radio continuum. The velocity coverage and resolution for H{\sc i} were 164 and 1.3 km s$^{-1}$, respectively. The typical noise fluctuations {were} $\sim$0.93 K at a velocity resolution of 1.3 km s$^{-1}$ for the H{\sc i} {and} $\sim$0.19 K for the radio continuum.

\subsection{Dust {Optical Depth and} Temperature}
{To derive the CO-to-H$_2$ conversion factor toward NGC~2359, we also used maps of the {dust optical depth} $\tau_\mathrm{353}$ at 353 GHz and the dust temperature $T_\mathrm{d}$ obtained with the $Planck$ and $IRAS$ satellites. These maps were derived by modified blackbody using 100 $\mu$m of IRIS and 353, 545, and 857 GHz observed with $Planck$ (for more detail, see \cite{2011A&A...536A..24P}).} The angular resolution of these datasets was $\sim$4$'$. We utilized the data release version R1.20 (see \cite{2014A&A...571A..11P}).

\section{Results} \label{sec:results}
\subsection{{Distributions of molecular and atomic clouds}}\label{sec:ngc2359}
{Figure \ref{moment} shows the results of $^{12}$CO($J$ = 3--2) line emission toward NGC~2359: (a) peak intensity, (b) moment 1, and (c) moment 2. The maps of moment 1 and moment 2 indicate the peak velocity and velocity dispersion, respectively. We confirmed that there are three molecular clouds at the velocities of $\sim$37, $\sim$54, and $\sim$67 km s$^{-1}$, which were found by previous CO studies (e.g., \cite{1981ApJ...243..184S,2001AJ....121.2664C,2001A&A...366..146R,2003A&A...411..465R}). The cloud at the velocity of $\sim$37 km s$^{-1}$ (hereafter referred to as ``37 km s$^{-1}$ CO cloud'') is located on the eastern side of the WR star with an elliptical shape. The cloud at the velocity of $\sim$54 km s$^{-1}$ (hereafter referred to as ``54 km s$^{-1}$ CO cloud'') consists of a bent filament and several small clumps with a size of 1 pc or less, some of which are overlapped with the southern edge of the 37 km s$^{-1}$ CO cloud. Almost all of the 37 and 54 km s$^{-1}$ CO clouds are located within 10~pc in projection from the WR star. On the other hand, the cloud at the velocity of $\sim$67 km s$^{-1}$ (hereafter referred to as ``67 km s$^{-1}$ CO cloud'') is distant 10 pc or larger away from the WR star, {showing} highly filamentary distribution elongated in the south--southeastern direction. We also found that the velocity width of the 54 km s$^{-1}$ CO cloud is clearly larger than that of the other CO clouds (e.g., \cite{2001A&A...366..146R}). The above observational trends are also seen in $^{12}$CO($J$ = 1--0) line emission but is noisy due to slightly higher noise fluctuation level. Hereafter, we mainly use the $^{12}$CO($J$ = 3--2) line emission data as a tracer of molecular clouds in NGC~2359.}

{Figure \ref{channel_aste} shows the contour maps of velocity channel distribution of $^{12}$CO($J$ = 3--2) toward NGC~2359 superposed on the radio continuum image. We find that the 54 km s$^{-1}$ CO cloud shows a good spatial correspondence with the radio continuum image (Figures \ref{channel_aste}e and \ref{channel_aste}f). More precisely, the radio continuum emission is enhanced toward the 54 km s$^{-1}$ CO cloud. On the other hand, there is no enhancement of radio continuum emission toward the 67 km s$^{-1}$ CO cloud, since the CO cloud is located on the outside of the radio shell boundary (Figure \ref{channel_aste}i). For the 37 km s$^{-1}$ CO cloud, we find a weak limb-brightening of radio continuum toward the southwestern edge of the CO cloud (Figure \ref{channel_aste}b).}

Figure \ref{channelmap} shows the $^{12}$CO($J$ = 3--2) and H{\sc i} channel maps in which the radio-continuum boundaries are superposed. The H{\sc i} distribution exhibits bright structures at $V_\mathrm{LSR}$ $\sim$54 (Figures \ref{channelmap}e--\ref{channelmap}g) and $\sim$63 km s$^{-1}$ (Figure \ref{channelmap}h), as previously found by \citet{1999AJ....118..948C}. The component at $V_\mathrm{LSR}$ $\sim$54 km s$^{-1}$ consists of two filamentary H{\sc i} clouds: one extends from east to southwest (hereafter referred to as the ``{S}outh H{\sc i} cloud''), and the other is elongated from north to northwest of the nebula (hereafter referred to as the ``{N}orth H{\sc i} cloud''). Both the North and South H{\sc i} clouds {are located along the edge of the radio continuum shell.} We also find that the South H{\sc i} cloud shows a good spatial correspondence not only with the radio shell boundary, but also with the 54 km s$^{-1}$ CO cloud. In Figure \ref{channelmap}h, the southern part of the radio boundary near {($\alpha_\mathrm{J2000}$, $\delta_\mathrm{J2000}$) $\sim$ ($07^{\mathrm{h}}18^{\mathrm{m}}42^{\mathrm{s}}$, $-13{^\circ}17\arcmin00\arcsec$) is} deformed along with the H{\sc i} {cloud} at $V_\mathrm{LSR}$ $\sim$63 km s$^{-1}$ {(hereafter referred to as the ``63 km s$^{-1}$ H{\sc i} cloud'')}.

{Figure \ref{hi_co_pv}(b) shows the position--velocity diagram for H{\sc i} and $^{12}$CO($J$ = 3--2) within the region shown by two lines in Figure \ref{hi_co_pv}(a). We newly find a clear evidence for an expanding shell both the H{\sc i} and CO with an expansion velocity of $\sim$4.5 km s$^{-1}$ (see dashed circle in Figure \ref{hi_co_pv}b). The center velocity of the expanding shell is roughly 51 km s$^{-1}$, which was derived the lowest intensity position of H{\sc i} within the ring-like structure in the position-velocity diagram. This value is roughly consistent with the {systemic} velocity of NGC~2359 which was estimated by optical spectroscopy (\cite{1983A&A...117..127G}). Since the spatial extent of expanding shell is also consistent with that of the radio continuum shell, the expanding shell was likely formed by strong feedback from the WR star. This is an alternative support for the shock and/or wind acceleration of the 54 km s$^{-1}$ CO cloud presented by \citet{2003A&A...411..465R}. We also note that an H{\sc i} component at $V_\mathrm{LSR}$ $\sim$41 km s$^{-1}$ appears to connect the 37 and 54 km s$^{-1}$ CO clouds. In addition to this, a diffuse H{\sc i} component at $V_\mathrm{LSR}$ $\sim$60 km s$^{-1}$ is also connecting the H{\sc i} clouds at the velocities of 54 and 63 km s$^{-1}$ (see the rectangles in Figure \ref{hi_co_pv}b).}

\subsection{Physical {properties of} CO {and H{\sc i}} clouds}\label{sec:phys}
To estimate the mass of {the CO clouds, $M_\mathrm{CO}$ }, we used the following equation:
\begin{eqnarray}
M_\mathrm{CO} = \mu m_{\mathrm{H}} \Omega d^2 \sum_{i}N_i(\mathrm{H}_2),
\label{eq6}
\end{eqnarray}
where $\mu = 2.8$ is mean molecular weight, taking into account a helium abundance of 20$\%$, $m_{\mathrm{H}}$ is the mass of the hydrogen atom, $\Omega$ is the solid angle of a square pixel, $d$ is the distance to the molecular cloud, and $N_i(\mathrm{H}_2)$ is the molecular hydrogen column density for each pixel. We used the following equation to derive $N(\mathrm{H}_2)$: 
\begin{eqnarray}
N(\mathrm{H_2}) = X(\mathrm{CO}) \cdot W(\mathrm{CO})
\label{eq2}
\end{eqnarray}
where {$X(\mathrm{CO}$) is the CO-to-H$_2$ conversion factor, and $W(\mathrm{CO})$ is the integrated intensity of $^{12}$CO($J$ = 1--0) line emission. In the present paper, we utilize $X(\mathrm{CO})$ = 1.9 $\times$ 10$^{20}$ cm$^{-2}$ (K km s$^{-1}$)$^{-1}$, which is derived in the Appendix.} The mass of {the} molecular cloud{s are $\sim$630 $M_\odot$ for the 37 km s$^{-1}$ CO cloud{;} $\sim$890 $M_\odot$ for the 54 km s$^{-1}$ CO cloud{;} and $\sim$240 $M_\odot$ for the 67 km s$^{-1}$ CO cloud} at the distance of 5 kpc. The maximum value{s} of $N(\mathrm{H}_2)$ for the 37 {and 67 km s$^{-1}$ CO clouds are found to be $\sim$3.1 and 1.1 $\times$ 10$^{21}$ cm$^{-2}$ respectively, whereas that of the 57 km s$^{-1}$ CO cloud is $\sim$7.8 $\times$ 10$^{21}$ cm$^{-2}$.}

{For the mass of H{\sc i} clouds $N(\mathrm{H}{\textsc{i}})$, we used the following equation:}
\begin{eqnarray}
M_\mathrm{HI} = m_{\mathrm{H}} \Omega d^2 \sum_{i}N_i(\mathrm{H}{\textsc{i}}).
\label{eq6}
\end{eqnarray}
{To derive $N(\mathrm{H}{\textsc{i}})$, it is thought to be considered that H{\sc i} clouds are optically thin (optical depth of H{\sc i} $\ll 1$). In this case, $N(\mathrm{H}{\textsc{i}})$ can be descried as below:}
\begin{eqnarray}
N(\mathrm{H}{\textsc{i}}) = 1.823 \times 10^{18}\phantom{0}W(\mathrm{H}{\textsc{i}}),
\label{eq5}
\end{eqnarray}
{where $W$(H{\sc i}) is the integrated intensity of H{\sc i}. On the other hand, \citet{2015ApJ...807L...4F} argued that 85\% of atomic hydrogen in the local interstellar H{\sc i} cloud is optically thick (optical depth of H{\sc i} $\sim$0.5--3) respect to the H{\sc i} 21 cm line. Subsequent detailed studies by \citet{2017ApJ...850...71F} demonstrated that the optical-depth corrected $N(\mathrm{H}{\textsc{i}})$ as a function of $W$(H{\sc i}) using the maps of dust optical depth at 353~GHz datasets assuming the nonlinear dust properties (more detailed information, see Appendix and equation \ref{eq1}). In the present study, we derive $N(\mathrm{H}{\textsc{i}})$ using the relation of \citet{2017ApJ...850...71F} as a function of $W$(H{\sc i}). The derived masses of H{\sc i} clouds are $\sim$130 $M_\odot$ for the North H{\sc i} cloud; $\sim$660 $M_\odot$ for the South H{\sc i} cloud; and $\sim$990 $M_\odot$ for the 67 km s$^{-1}$ H{\sc i} cloud. The masses of these clouds are 4--7 times higher than the results in the optically thin cases.} We summarize the physical properties of the CO and H{\sc i} clouds in Table \ref{table:extramath}, {and their radiation temperature, peak velocities, and velocity widths are derived by fitting a single Gaussian function.}

We investigate the physical condition of the three CO clouds using the ASTE $^{12}$CO($J$ = 3--2) and Nobeyama 45-m $^{12}$CO($J$ = 1--0) datasets. Figure \ref{ratio} shows the intensity ratio maps of CO $J$ = 3--2 / 1--0, hereafter {referred} to as ``$R_\mathrm{3-2/1-0}$'' toward the three CO clouds. For the 54 km s$^{-1}$ CO cloud, we find high-intensity ratio of $R_\mathrm{3-2/1-0} > 1.0$ from the entire region, which nicely traces the radio continuum shell. By contrast, the 67 km s$^{-1}$ CO cloud shows the lowest intensity ratio of $R_\mathrm{3-2/1-0} \lesssim 0.3$. For the 37 km s$^{-1}$ CO cloud, we find slightly higher value of $R_\mathrm{3-2/1-0}$ $\sim$0.7 {near the position of peak A.}

{To reveal detailed physical conditions of the CO clouds, we performed a Large Velocity Gradient analysis (LVG; e.g., \cite{1974ApJ...189..441G}; {\cite{1974ApJ...187L..67S}}) toward the CO peaks A, B, and C using the ASTE $^{12}$CO($J$ = 3--2) and Nobeyama 45-m $^{12}$CO and $^{13}$CO($J$ = 1--0) datasets. We adopt the velocity gradient $dv/dr = 0.5$ km s$^{-1}$ pc$^{-1}$ for peaks A and C; and 2.5 km s$^{-1}$ pc$^{-1}$ for peak B, where $dv$ is the full-width-half-maximum (FWHM) line width of CO and $dr$ is estimated as an effective diameter of the area whose integrated intensity of CO exceeds half of the peak. We also assumed the abundance ratios of [$^{12}$CO/H$_2$] = $5 \times 10^{-5}$ (\cite{1987ApJ...315..621B}) and [$^{12}$CO/$^{13}$CO] = 75 (\cite{2004dimg.conf..253G}). Finally, we adopt $X/(dv/dr$) = 1 $\times$ 10$^{-4}$ (km s$^{-1}$ pc$^{-1}$)$^{-1}$ for peaks A and C; and 2 $\times$ 10$^{-5}$ (km s$^{-1}$ pc$^{-1}$)$^{-1}$ for peak B.}

{To solve the number density of molecular hydrogen $n(\mathrm{H_2})$ and kinetic temperature $T_\mathrm{kin}$, we calculate $\chi^2$ defined as}
\begin{eqnarray}
\chi^2 = \sum[(R_\mathrm{obs} - R_\mathrm{model})^2 / \sigma],
\label{eq00}
\end{eqnarray}
{where $R_\mathrm{obs}$ is the observed intensity ratio of different excitation lines or isotopes, $R_\mathrm{model}$ is the line ratio of the LVG model, and $\sigma$ is the standard deviation for $R_\mathrm{obs}$. The error of $R_\mathrm{obs}$ is estimated by considering the noise fluctuation and the intensity calibration error. We adopt the error of the intensity calibration from the antenna temperature $T_A^*$ to the main beam temperature $T_\mathrm{MB}$ is 10\%.}

{Figure \ref{lvg} shows the CO spectra and LVG results toward the positions of peaks A, B, and C. For the CO spectra, we used the velocity ranges of 1.2 km s$^{-1}$ widths for the LVG analysis, which are shown by the shaded areas in Figures \ref{lvg}(a--c). For the LVG results, we have estimated the best values of $n(\mathrm{H_2})$ and $T_\mathrm{kin}$ for each peak as shown in crosses, which are the lowest points of $\chi^2$. We can also reject the each area outside of the black contours at the 95\% confidence level, corresponding to $\chi^2 = 3.84$ (d.o.f. = 1). We find the highest density $n(\mathrm{H_2})$ and temperature $T_\mathrm{kin}$ toward peak B in the 54 km s$^{-1}$ CO cloud: $n(\mathrm{H_2}) = 2.6 \times 10^{4}$ cm$^{-3}$ and $T_\mathrm{kin} = 63$ K. By contrast, the lowest values of them are found toward peak C in the 67 km s$^{-1}$ CO cloud: $n(\mathrm{H_2}) = 600$ cm$^{-3}$ and $T_\mathrm{kin} = 6$ K, which are typical values of quiescent molecular clouds. For peak A in the 37 km s$^{-1}$ CO cloud, we find slightly higher temperature of $T_\mathrm{kin} = 17$ K, whereas the density is not really $n(\mathrm{H_2}) = 800$ cm$^{-3}$.}

\section{Discussion} \label{sec:discuss}
\subsection{{Molecular and Atomic clouds associated with the Wolf-Rayet nebula NGC~2359}}\label{association}
{In the present study, we have confirmed the presence of three molecular clouds (37, 54, and 67 km s$^{-1}$ CO clouds) and three H{\sc i} clouds (North, South, and 63 km s$^{-1}$ H{\sc i} clouds) toward NGC~2359 using the ASTE $^{12}$CO($J$ = 3--2), Nobeyama 45-m CO($J$ = 1--0), and the VLA H{\sc i} datasets as also mentioned by previous studies (e.g., \cite{1981ApJ...243..184S,1999AJ....118..948C,2001AJ....121.2664C,2001A&A...366..146R,2003A&A...411..465R}). These previous CO/H{\sc i} studies argued that four of them---the 54 km s$^{-1}$ CO cloud, North, South, and 63 km s$^{-1}$ H{\sc i} clouds---are probably associated with the WR nebula NGC~2359. However, it is still under debate whether all the CO/H{\sc i} clouds are associated with NGC~2359 or not. In this section, we shall discuss which clouds are physically associated with NGC~2359.}

{First of all, we argue that the 54 km s$^{-1}$ CO cloud and the North/South H{\sc i} clouds are definitely interacting with NGC~2359. In the present study, we newly found that the 54 km s$^{-1}$ CO cloud is limb-brightened in radio continuum (Figures \ref{channel_aste}e and \ref{channel_aste}f), suggesting that the surface of molecular clouds are partially ionized by strong UV radiation and/or stellar winds from the WR star. The high-kinematic temperature $T_\mathrm{kin}$ of 63~K could be also interpreted as a natural result of heating due to the strong feedback from the WR star (Figure \ref{lvg}e). For the North/South H{\sc i} clouds, we found a good spatial correspondence with the boundary of radio continuum shell (Figure \ref{channelmap}). This suggests that the part of H{\sc i} clouds near the WR star have been ionized and the produced ionized gas emits the bright radio continuum.}

\begin{figure*}
\begin{center}
\includegraphics[width=\linewidth]{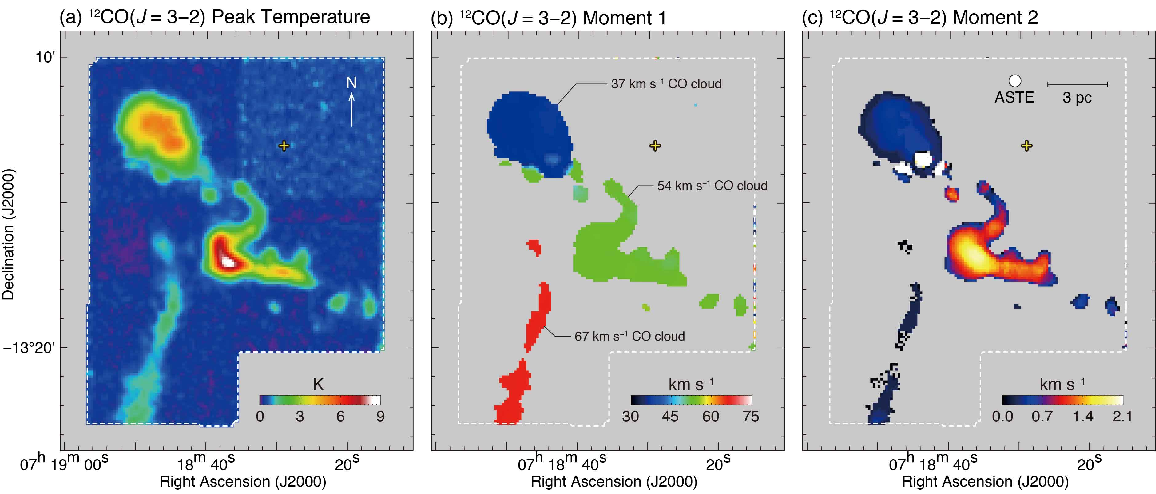}
\end{center}
\caption{{Results of $^{12}$CO($J$ = 3--2) observations toward NGC~2359. Maps of (a) peak intensity, (b) moment 1, and (c) moment 2. The dashed polygon indicates the observed region. The beam size and scale bar are also shown in right top corner of Figure \ref{moment}(c). The 37, 54, and 67 km s$^{-1}$ CO clouds are also indicated in Figure \ref{moment}(b). The crosses indicate the position of the WR star.}}
\label{moment}
\end{figure*}

\begin{figure*}
\begin{center}
\includegraphics[width=\linewidth]{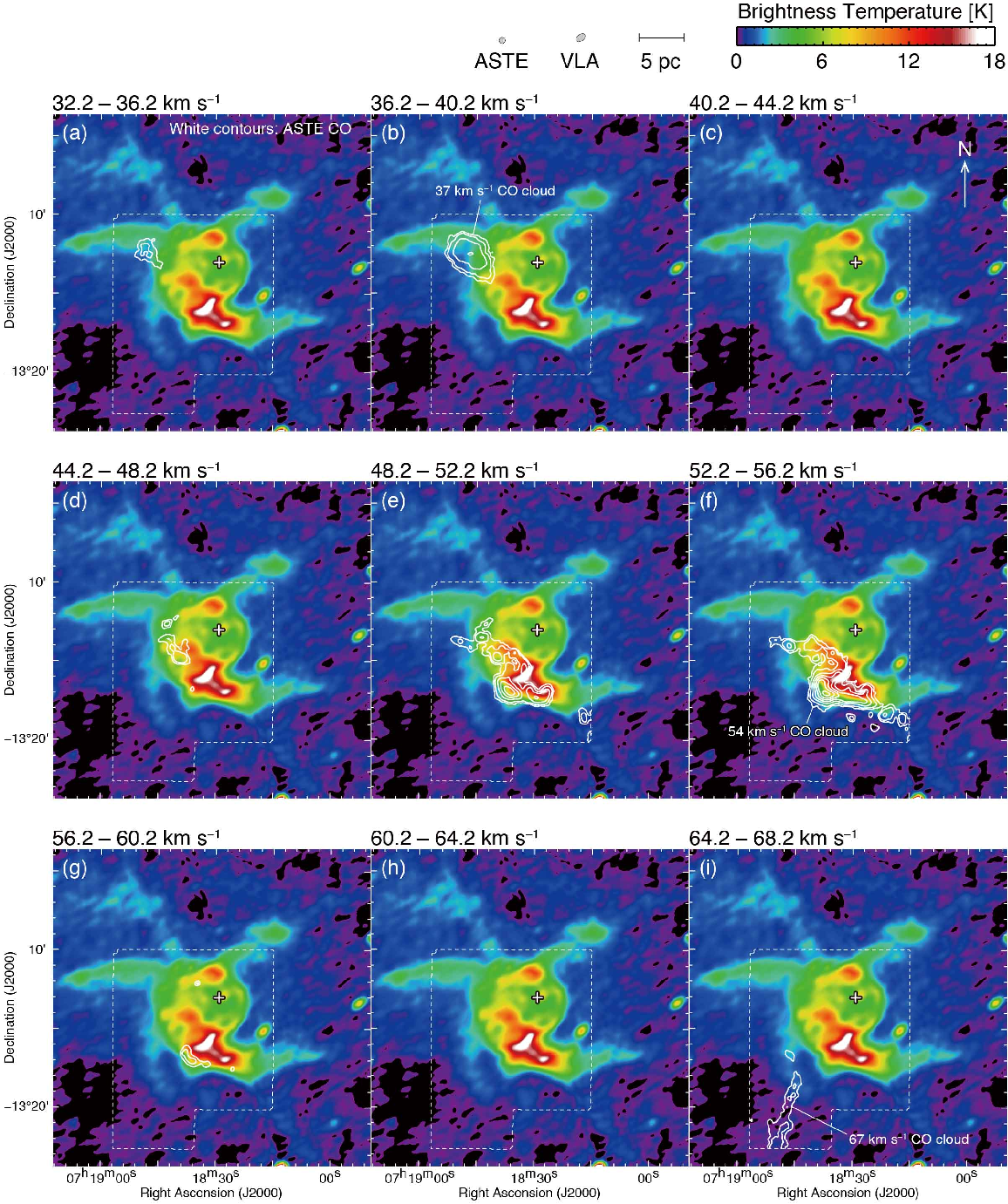}
\end{center}
\caption{{Velocity channel maps of $^{12}$CO($J$ = 3--2) brightness temperature ($white$ $contours$) superposed on the radio continuum image. Each panel of CO contours shows intensity distributions averaged every 4.0 km s$^{-1}$ in a velocity range from 32.2 to 68.2 km s$^{-1}$. The contour levels of CO are 0.20, 0.36, 0.84, 1.65, 2.78, 4.23, and 6.00 K. The beam size and scale bar are also shown in the top right corner. The crosses indicate the position of the WR star.}}
\label{channel_aste}
\end{figure*}

\begin{figure*}
\begin{center}
\includegraphics[width=\linewidth]{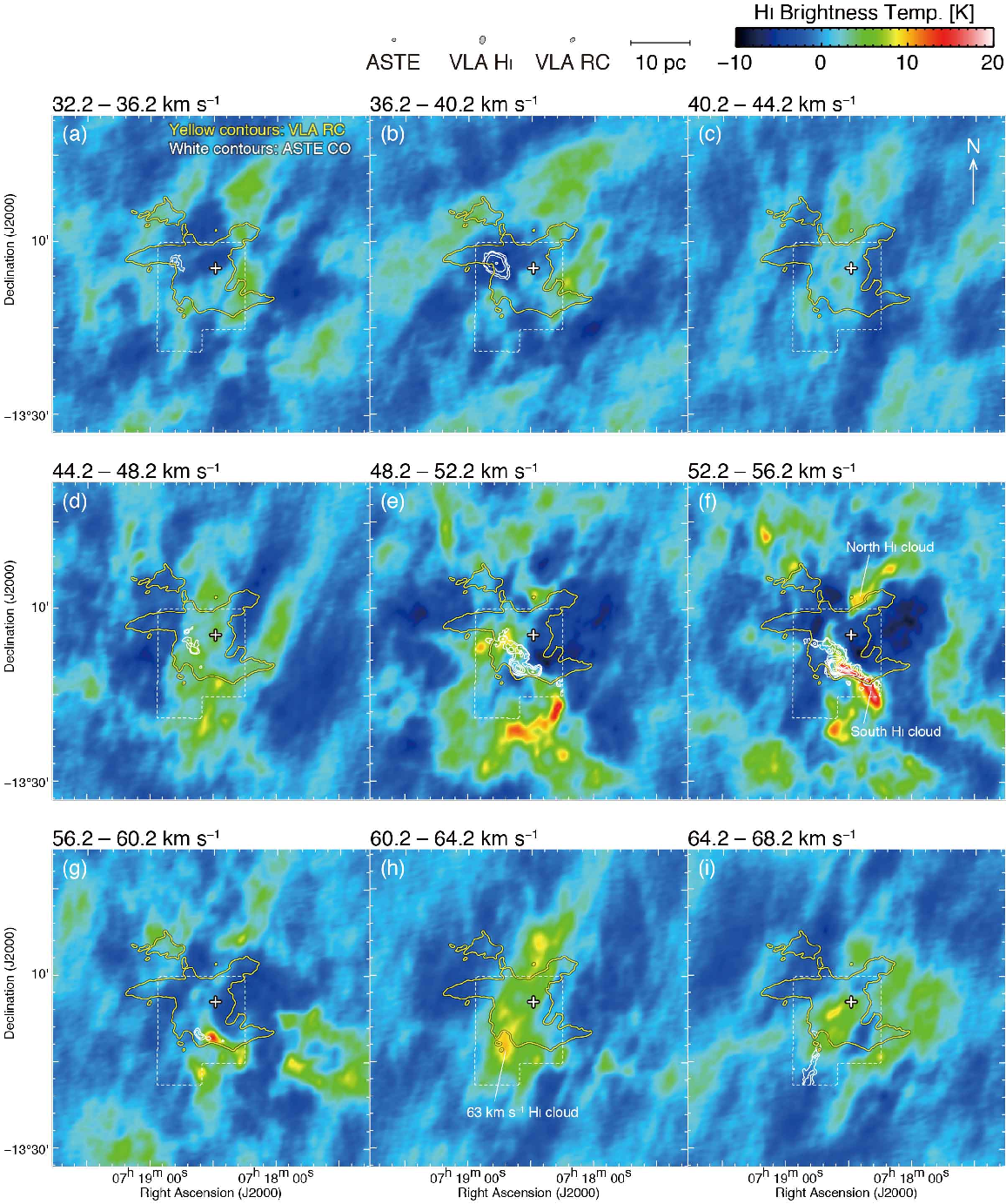}
\end{center}
\caption{Velocity channel maps of H{\sc i} ($color$ $image$) and $^{12}$CO($J$ = 3--2) brightness temperature ($white$ $contours$). Superposed yellow contours indicate the radio continuum boundary of 1.5 K. The velocity range and intervals are the same as shown in Figure \ref{channel_aste}. The beam size and scale bar are also shown in the top right corner. {The crosses indicate the position of the WR star}.}
\label{channelmap}
\end{figure*}

\begin{figure*}
\begin{center}
\includegraphics[width=\linewidth]{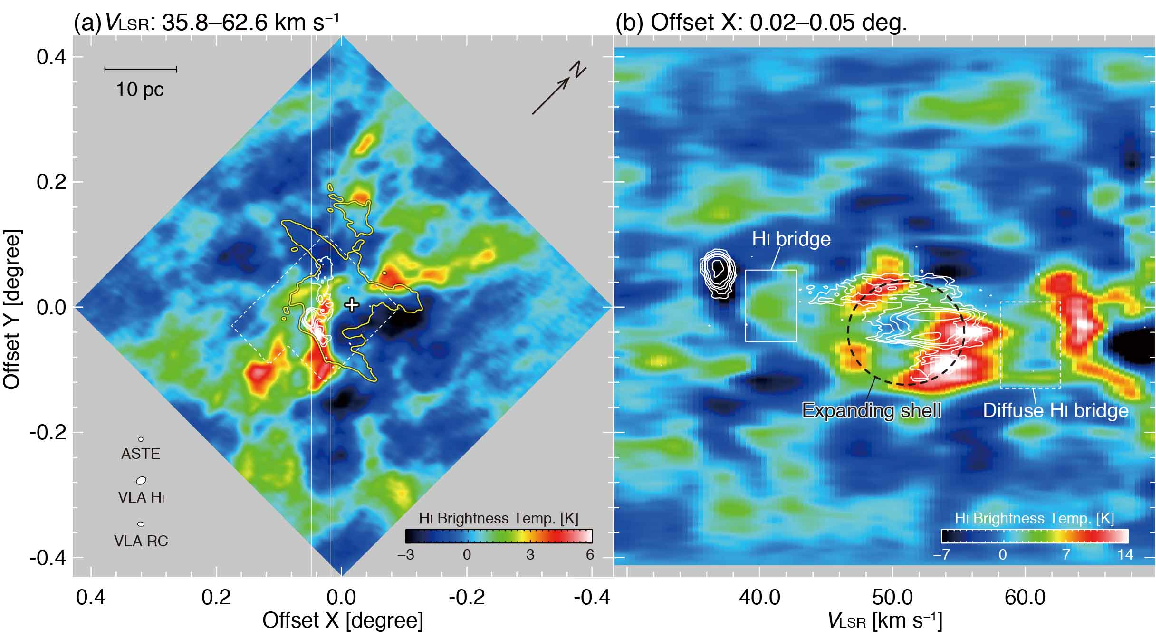}
\end{center}
\caption{{(a) Integrated intensity map of H{\sc i} ({\it color image}) superposed on the $^{12}$CO($J$ = 3--2) ({\it white contours}) and the radio continuum boundary of 1.5 K ({\it yellow contours}). The map is rotated 45 degrees clockwise with respect to north. The integrated velocity range{s} of H{\sc i} and CO {are} from 35.8 {to} 62.6 km s$^{-1}$. The contour levels of CO {are} from 0.15, 0.3, 0.6, 0.9, 1.5, 3.0, 4.5, and 6.0 K. The cross symbol indicates the position of the WR star. The beam size and scale bar are also shown in the bottom-left corner. (b) Position-velocity diagram of the H{\sc i} superposed on the $^{12}$CO($J$ = 3--2) contours. The integration range of Offset-X is from 0\fdg02 to 0\fdg05. The contour levels of CO are 0.15, 0.3, 0.6, 0.9, 1.5, 3.0, 6.0, 9.0, and 12.0 K. The dashed circle and white boxes indicate the expanding shells of CO/H{\sc i} and H{\sc i} bridge, respectively (see the text).}}
\label{hi_co_pv}
\end{figure*}

\begin{figure*}
\vspace*{1cm}
\begin{center}
\includegraphics[width=\linewidth]{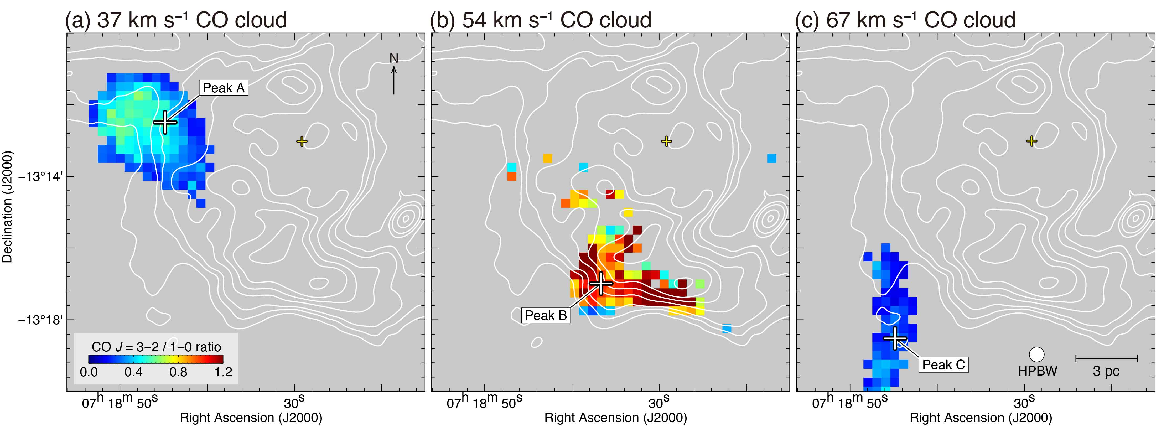}
\end{center}
\caption{Maps of $^{12}$CO $J$ = 3--2 / 1--0 ratio toward the (a) 37 km s$^{-1}$ CO cloud {,} (b) 54 km s$^{-1}$ CO cloud{,} and the 67 km s$^{-1}$ CO cloud. {The gray shaded areas represent that the $^{12}$CO($J$ = 1--0) and/or $^{12}$CO($J$ = 3--2) data show the low significance of $\sim$3$\sigma$ or lower. Superposed contours indicate the radio continuum as shown in Figure \ref{top}. The integrated velocity range is from 35.8 to 37.8 km s$^{-1}$ for the 37 km s$^{-1}$ CO cloud; from 42.6 to 57.0 km s$^{-1}$ for the 54 km s$^{-1}$ CO cloud; and from 65.4 to 67.4 km s$^{-1}$ for the 67 km s$^{-1}$ CO cloud. Each plot is the same intensity scale as shown in the bottom left corner in Figure \ref{ratio}(a). The beam size and scale bar are also shown in the bottom right corner in Figure \ref{ratio}(c). The yellow crosses indicate the position of the WR star. The CO intensity peaks A--C for the LVG analysis are also indicated by white crosses (see the text).}}
\label{ratio}
\vspace*{1cm}
\end{figure*}

\begin{figure*}
\begin{center}
\includegraphics[width=\linewidth]{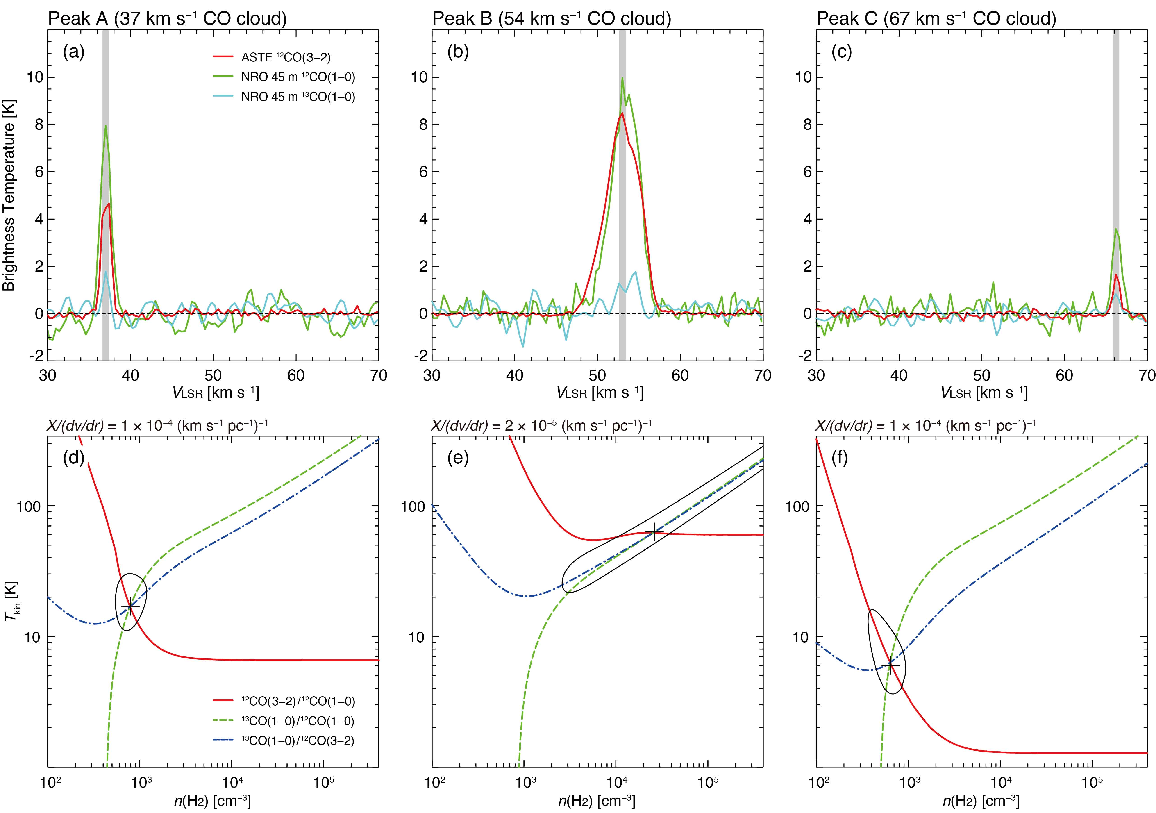}
\end{center}
\caption{{({\it{upper panels}}) CO spectra toward (a) peak A, (b) peak B, and (c) peak C. The red, green, and blue represent $^{12}$CO($J$ = 3--2), $^{12}$CO($J$ = 1--0), and $^{13}$CO($J$ = 1--0) line emission, respectively. (b) LVG results on the number density of molecular hydrogen $n$(H$_2$) and kinematic temperature $T_\mathrm{kin}$ plane for the (d) 37 km s$^{-1}$ CO cloud, (e) 54 km s$^{-1}$ CO cloud, and (f) 67 km s$^{-1}$ CO cloud. Each line indicates the intensity ratios of $^{12}$CO $J$ = 3--2 / 1--0 (red), $^{13}$CO $J$ = 1--0 / $^{12}$CO $J$ = 1--0 (green), and $^{13}$CO $J$ = 1--0 / $^{12}$CO $J$ = 3--2 (blue). The best values of $n$(H$_2$) and $T_\mathrm{kin}$ plane are also shown by crosses. Black contours mean the 95\% confidence level of a $\chi^2$ distribution. The values of $X/(dv/dr)$ are also shown in the top left corner in each panel.}}
\label{lvg}
\end{figure*}

{The velocity structures of CO/{H\sc i} emission also provide an alternative support for the association. According to \citet{2003A&A...411..465R}, the 54 km s$^{-1}$ CO cloud shows three different velocity components: low velocity component ({\it{lvc}}) at velocities from 52 to 57 km s$^{-1}$, intermediate velocity component ({\it{ivc}}) at velocities from 48 to 52 km s$^{-1}$, and high velocity component ({\it{hvc}}) at velocities lower than 48 km s$^{-1}$. The authors concluded that the multiple layers of {\it{lvc--ivc--hvc}} in the 54 km s$^{-1}$ CO cloud were produced by different evolutive episodes of the progenitor of the WR star, such as the expanding stellar winds or mass ejections during the earlier luminous blue variable (LBV) phase and/or current WR stage. In our analysis using $^{12}$CO($J$ = 3--2) and H{\sc i}, we confirmed that the three velocity components are part of the expanding CO/H{\sc i} shell originated by the strong feedback from the WR star and its progenitor (Figure \ref{hi_co_pv}b). Furthermore, the wider line widths of the 54 km s$^{-1}$ CO cloud and South H{\sc i} cloud are also consistent with the acceleration by stellar winds (see Figure \ref{moment}c and Table \ref{table:extramath}). We therefore conclude that the three CO/{H\sc i} clouds at 54 km s$^{-1}$ are likely associated with the WR Nebula NGC~2359.}

{Next, we focus on the 63 km s$^{-1}$ H{\sc i} cloud. As for the morphological aspects, we found the deformation of radio shell along the intensity peak of the 63 km s$^{-1}$ H{\sc i} cloud (Figure \ref{channelmap}h). It is likely that the dense H{\sc i} cloud has been survived by shock erosion and/or UV radiation, and hence the ionized region is extended to avoid the cloud. We also note that the 63 km s$^{-1}$ H{\sc i} cloud is connected to the expanding shell by the diffuse bridging feature of H{\sc i} in the position--velocity diagram (see white dashed box in Figure \ref{hi_co_pv}b). Although the H{\sc i} cloud is not located within the expanding shell in the velocity space, the bridging feature provides an alternative support for the physical relation between the 63 km s$^{-1}$ H{\sc i} cloud and NGC~2359. The origin of velocity difference between the expanding shell and the H{\sc i} cloud will be discussed in Section \ref{sec:feedback}. In any case, we argue that the 63 km s$^{-1}$ H{\sc i} cloud is also associated with NGC~2359 as previously mentioned by \citet{1999AJ....118..948C}.}

{For the 67 km s$^{-1}$ CO cloud, however, there is no strong sign of the interaction with NGC~2359. Figure \ref{channel_aste}(i) shows that the CO cloud is located on the outside of the bright radio shell. The northern edge of the CO cloud is possibly in contact with the diffuse radio emission, but is no significant enhancement of radio flux and line width of $^{12}$CO($J$ = 3--2) line emission (see Figures \ref{moment}c and \ref{channel_aste}i). Since the kinematic temperature is also below 10 K, we conclude that the 67 km s$^{-1}$ CO cloud seems to be unconnected with NGC~2359.}

{Finally we discuss the 37 km s$^{-1}$ CO cloud, which is still under debate whether the cloud is associated with the nebula or not. \citet{1999AJ....118..948C} argued that it is not clear if the CO cloud at 37 km s$^{-1}$ is associated with NGC~2359 from the present datasets. On the other hand, \citet{2001A&A...366..146R} concluded that the 37 km s$^{-1}$ CO cloud is not associated with the nebula from the view points of narrow line width of up to 2 km s$^{-1}$ and no sign of morphological or kinematical disturbance from the WR star and its optical nebula. We also confirmed the narrow line width of CO as shown in Figure \ref{moment}(c). By contrast, we newly find two hints that support the physical association between the cloud and nebula: the morphological aspects and physical condition of the cloud. By comparative studies of CO and radio continuum image, we found the weak enhancement of radio continuum at the southern edge of the 37 km s$^{-1}$ CO cloud (Figure \ref{channel_aste}b). This is a possible evidence of ionization on the surface of the cloud similar to the 54 km s$^{-1}$ CO cloud. In fact, the kinematic temperature of the 37 km s$^{-1}$ CO cloud ($T_\mathrm{kin} \sim$17 K) is three times higher than that of the 67 km s$^{-1}$ CO cloud ($T_\mathrm{kin} \sim$6 K), indicating that the shock or UV heating likely occurred. Since the kinematic temperature of the 37 km s$^{-1}$ CO cloud is not as high as that of the 54 km s$^{-1}$ CO cloud ($T_\mathrm{kin} \sim$63 K), implying that the 37 km s$^{-1}$ CO cloud is probably associated with NGC~2359, but is slightly distant from the WR star compared to the 54 km s$^{-1}$ CO cloud.}

\begin{figure*}
\begin{center}
\includegraphics[width=\linewidth]{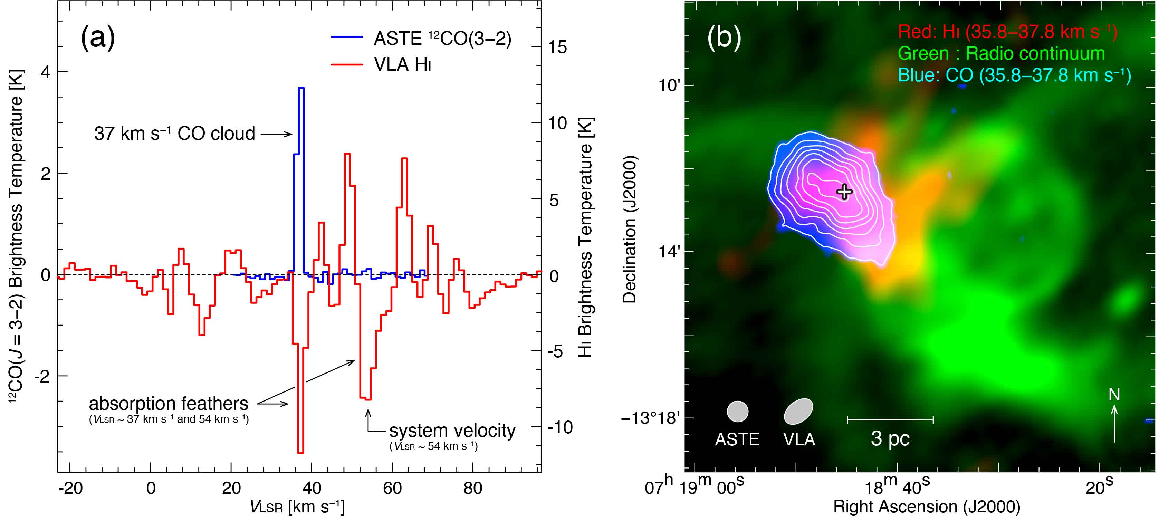}
\end{center}
\caption{{(a) $^{12}$CO($J$ = 3--2) and H{\sc i} spectra toward the 37 km s$^{-1}$ CO cloud. The spectra were extracted by the position of Peak A (white cross in Figure \ref{hiabs}b). (b) Three color image of the 37 km s$^{-1}$ cloud. The red, green, and blue represent the H{\sc i}, radio continuum, and $^{12}$CO($J$ = 3--2). The integration velocity ranges of CO and H{\sc i} are from 35.8 to 37.8 km s$^{-1}$ which are the same velocities of the 37 km s$^{-1}$ CO cloud. The white contours indicate the integrated intensity of CO, whose contour levels are 1, 2, 3, 4, 5, 6, and 7 K km s$^{-1}$. For the H{\sc i}, the intensity scale is negative. The beam size and scale bar are also shown in the bottom left corner.}}
\label{hiabs}
\end{figure*}

{We also argue that the 37 and 54 km s$^{-1}$ CO clouds appear to be connected by the H{\sc i} bridging feature in the position--velocity diagram (see white solid box in Figure \ref{hi_co_pv}b). Although the H{\sc i} emission at the velocity of 37 km s$^{-1}$ CO cloud is missing at first glance, we find ``an H{\sc i} cloud'' at the velocity which is physically connected to the H{\sc i} bridge, 37 and 54 km s$^{-1}$ CO clouds. Figure \ref{hiabs}(a) shows the $^{12}$CO($J$ = 3--2) and H{\sc i} spectra toward the 37 km s$^{-1}$ CO cloud. We find two absorption lines of H{\sc i} at the velocities of 37 and 54 km s$^{-1}$. The former shows a good agreement with the central velocity of the 37 km s$^{-1}$ CO cloud. The latter corresponds to the {systemic} velocity of NGC~2359. These suggest that the H{\sc i} clouds at the velocities of 37 and 54 km s$^{-1}$ are observed as absorption lines due to the strong radio continuum emission from NGC~2359 as the background source. In this case, we can find a good spatial correspondence among the 37 km s$^{-1}$ CO cloud, absorption line of H{\sc i}, and the radio continuum emission. Figure \ref{hiabs}(b) shows the three color image of the 37 km s$^{-1}$ CO cloud (blue), absorption line of H{\sc i} (red), and the radio continuum emission (green). For the H{\sc i}, the intensity scale is negative. We can clearly see a good spatial correspondence among them, and hence the H{\sc i} cloud is associated with the 37 km s$^{-1}$ CO cloud. Since the H{\sc i} bridge is extended to the velocity of 37 km s$^{-1}$, we propose that the 37 km s$^{-1}$ CO cloud is physically connected with the 54 km s$^{-1}$ CO cloud by the H{\sc i} bridging feature. On the basis of these results we conclude that the the 37 km s$^{-1}$ CO cloud is also likely interacting with the WR star and its nebula. To reliable confirm the cloud association of the 37 km s$^{-1}$ CO cloud, observations of photodissociation region tracers such as [C{\sc i}] and [C{\sc ii}] line emission using ALMA and SOFIA are useful.}

\begin{figure}
\begin{center}
\includegraphics[width=140mm]{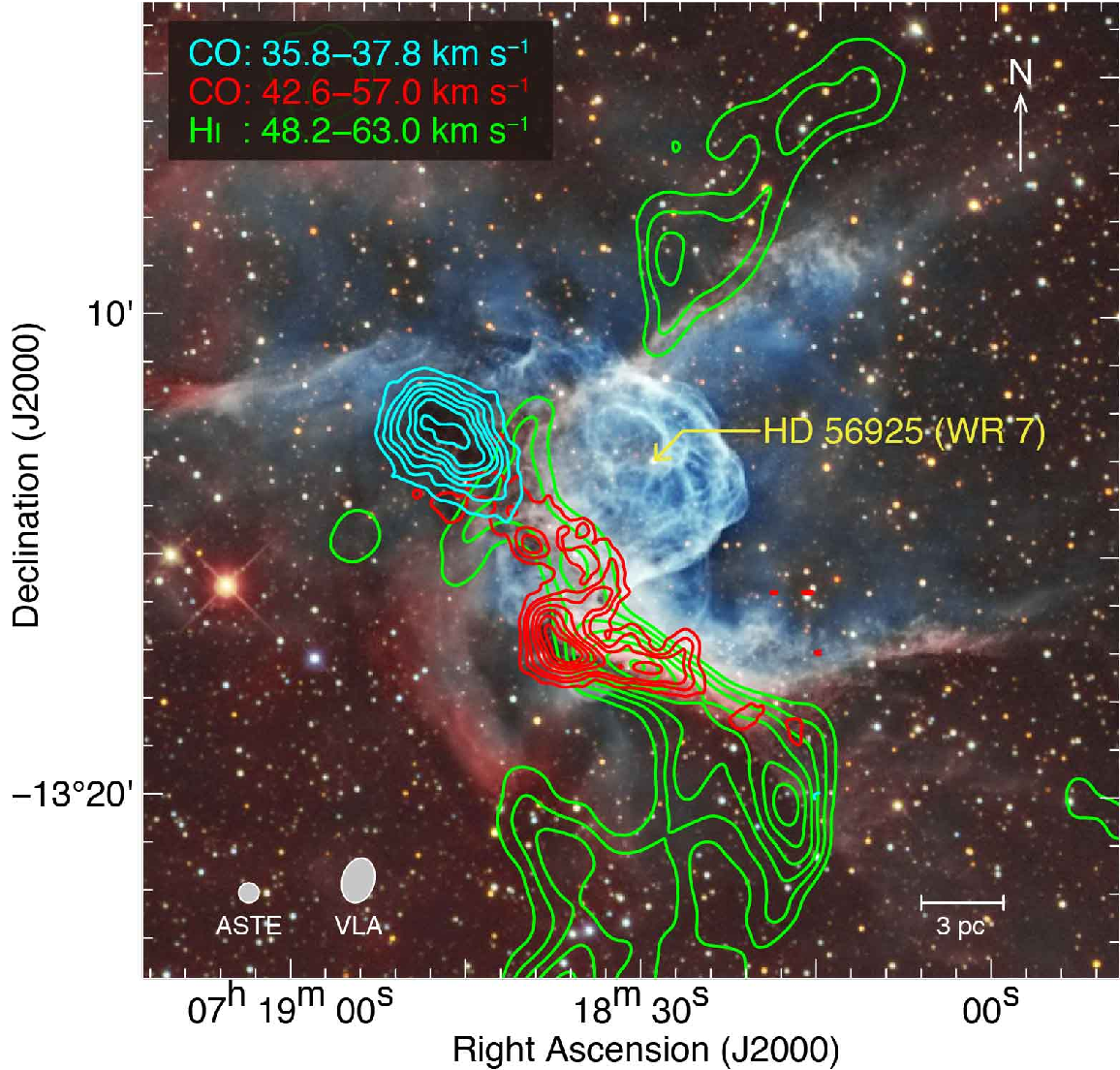}
\end{center}
\caption{{Optical image as shown in Figure \ref{top}, superposed on the $^{12}$CO($J$ = 3--2) contours ({\it blue and red}) and H{\sc i} contours ({\it green}). The integrated velocity range is from 35.8 to 37.8 km s$^{-1}$ for the 37 km s$^{-1}$ CO cloud; from 42.6 to 57.0 km s$^{-1}$ for the 54 km s$^{-1}$ CO cloud; and from 48.2 to 63.0 km s$^{-1}$ for the H{\sc i} cloud. The contour levels are 1, 2, 3, 4, 5, 6, and 7 K km s$^{-1}$ for the 37 km s$^{-1}$ CO cloud; 3, 6, 9, 15, 21, 27, and 33 K km s$^{-1}$ for 54 km s$^{-1}$ CO cloud; and 40, 60, 80, 100, 120, 140, and 160 K km s$^{-1}$ for the H{\sc i} cloud. The beam size and scale bar are also shown in the bottom right and left corners, respectively.}}
\label{overview}
\end{figure}

{To summarize, we conclude that the 37 and 54 km s$^{-1}$ CO clouds, and three H{\sc i} clouds at the velocities of 54 and 63 km s$^{-1}$ are physically associated with the WR nebula NGC~2359. Figure \ref{overview} shows the optical image superposed on the CO and H{\sc i} clouds associated with NGC~2359.} We find that the spatial distribution of the {37 km s$^{-1}$} CO cloud (blue contours) is complementary to that of the {54 km s$^{-1}$} CO cloud (red contours). In addition to this, both the South and North H{\sc i} clouds nicely trace not only the boundary of the radio shell but also the boundary of the optical filaments. In particular, the South H{\sc i} cloud is located along the {54 km s$^{-1}$} CO cloud up to the boundary of the {37 km s$^{-1}$} CO cloud. It is worth noting that the {37 and 54 km s$^{-1}$} CO clouds appear projected onto the optical dark lane, suggesting that both the CO clouds lie inside or in front of the nebula. These trends are consistent with the spatial comparison with the {optical emission} and previous CO studies (e.g., \cite{2001AJ....121.2664C,2001A&A...366..146R}, \yearcite{2003A&A...411..465R}).

\subsection{{Acceleration of CO/H{\sc i} clouds by strong stellar winds from the WR star and its progenitor}}\label{sec:feedback}
{In Section \ref{association}, we demonstrated that the 37 and 54 km s$^{-1}$ CO clouds, and three H{\sc i} clouds at the velocities of 54 and 63 km s$^{-1}$ are likely associated with NGC~2359. This means that some of clouds have different center velocities from the {systemic} velocity of the NGC~2359 system at $\sim$54 km s$^{-1}$ even though the all clouds are located at the roughly same distance from us. In this section, we discuss whether the velocity deferences of the CO/H{\sc i} can be explained by stellar feedback from the WR star and its progenitor.}

{We first argue that the velocity structures of three CO/H{\sc i} clouds at the velocities of $\sim$54 km s$^{-1}$ (54 km s$^{-1}$ CO cloud and the North/South H{\sc i} clouds) can be explained by acceleration due to strong stellar winds from the WR star and its progenitor. \citet{2003A&A...411..465R} considered two different models of stellar evolution: the progenitor mass of the WR star is 35 or 60 $M_\odot$ \citep{1994A&A...290..819L,1996A&A...305..229G}. In the present paper, we use the latter case because the higher mass case including the LBV phase is more compatible with the observational results in CO \citep{2003A&A...411..465R}. According to the model shown in \citet{1994A&A...290..819L}, the total released momentum of stellar winds is $\sim1.1 \times 10^5$ $M_\odot$ km s$^{-1}$. By considering the solid angle of each cloud with respect to the ionized cavity with a radius of $5'$, the transported wind momentum to the CO/H{\sc i} clouds is estimated to be $\sim4.5\times 10^3$ $M_\odot$ km s$^{-1}$ for the 54 km s$^{-1}$ CO cloud; $\sim3.1\times 10^4$ $M_\odot$ km s$^{-1}$ for the North H{\sc i} cloud; and $\sim7.6\times 10^4$ $M_\odot$ km s$^{-1}$ for the South H{\sc i} cloud. On the other hand, the CO/H{\sc i} clouds at the velocities of $\sim$54 km s$^{-1}$ show the expanding motion with the expansion velocity of $\sim$4.5 km s$^{-1}$ (see Figure \ref{hi_co_pv}b and Section \ref{sec:ngc2359}). We finally calculated the cloud momentum to be $\sim4.0\times 10^3$ $M_\odot$ km s$^{-1}$ for the 54 km s$^{-1}$ CO cloud;  $\sim5.7\times 10^2$ $M_\odot$ km s$^{-1}$ for the North H{\sc i} cloud; and $\sim3.0\times 10^3$ $M_\odot$ km s$^{-1}$ for the South H{\sc i} cloud. Since the cloud momentum is less than the wind momentum, the expanding motion of the CO/H{\sc i} clouds at velocities of $\sim$54 km s$^{-1}$ can be explained by the stellar wind acceleration.}

{The kinematics of the 63 km s$^{-1}$ H{\sc i} cloud can be also understood considering the stellar feedback in the same manner. The center velocity of 63 km s$^{-1}$ H{\sc i} cloud has an offset of $\sim$10 km s$^{-1}$ from the {systemic} velocity at $\sim$54 km s$^{-1}$. The cloud momentum of the 63 km s$^{-1}$ H{\sc i} cloud is estimated to be $\sim9.9\times 10^3$ $M_\odot$ km s$^{-1}$ if the velocity difference is originated by the wind acceleration, whereas the wind momentum is derived to $\sim7.6\times 10^4$ $M_\odot$ km s$^{-1}$. Therefore, the velocity difference of the 63 km s$^{-1}$ H{\sc i} cloud is also explained by the stellar wind acceleration. We note that the velocity difference is within a range of expansion velocity of ionized gas about $\pm$30 km s$^{-1}$ respect to the {systemic} velocity of NGC~2359 \citep{1983A&A...117..127G}.}

{Finally we argue that the kinematics of 37 km s$^{-1}$ CO cloud, however, cannot be accelerated only by the stellar winds from the WR star and its progenitor. Following the same manner, we estimate the cloud momentum of the 37 km s$^{-1}$ CO cloud to be $\sim1.1\times 10^4$ $M_\odot$ km s$^{-1}$ considering the velocity difference of $\sim$17 km s$^{-1}$ (see Table \ref{table:extramath}). On the other hand, the wind momentum for the CO cloud is estimated to $\sim5.8\times 10^3$ $M_\odot$ km s$^{-1}$, which is roughly twice less than the cloud momentum. Although the model parameters for wind momentum contain uncertainty, it is difficult to explain the velocity difference which was cased only by the stellar feedback.}

\subsection{{An alternative idea: isolated high-mass star formation triggered by a cloud-cloud collision}}\label{sec:ccc}
{In Section \ref{sec:feedback}, we discussed that the kinematics of the CO/H{\sc i} clouds can be explained by the wind acceleration of WR star and its progenitor, except for the 37 km s$^{-1}$ CO cloud. This indicates that the 37 km s$^{-1}$ CO cloud came from outside of the NGC~2359 system with the velocity difference of $\sim$17 km s$^{-1}$ or higher, and then interacted with the 54 km s$^{-1}$ CO cloud (and part of the North/South H{\sc i} clouds) because of the physical connection of the two CO clouds in the velocity space, and its association of the nebula (see Figure \ref{hiabs}b and Section \ref{association}). As one of the explanations of the observational results, we here propose a possible scenario that the progenitor of WR star was formed by collisions between the 37 and 54 km s$^{-1}$ CO (and H{\sc i}) clouds.}

{The velocity structures of the 37 and 54 km s$^{-1}$ CO clouds are compatible with the scenario of triggered star formation by the cloud-cloud collisions. According to \citet{2018ApJ...859..166F}, sites of high-mass star formation triggered by cloud collisions show that the two colliding clouds generally have a supersonic velocity separation at least a few km s$^{-1}$. Therefore, the large velocity separation of $\sim$17 km s$^{-1}$ of the 37 and 54 km s$^{-1}$ CO clouds satisfies the requirement. In fact, similar large velocity separation of $\sim$20 km s$^{-1}$ between the two colliding clouds is seen in the Galactic H{\sc ii} region RCW~120 which contains an isolated O-type star \citep{2015ApJ...806....7T}. The theoretical studies also mentioned the importance of the supersonic velocity separation. The numerical simulation by \citet{2013ApJ...774L..31I} confirmed that the effective Jeans mass in the shocked layer is proportional to the third power of the effective sound speed. Here, the effective sound speed is written by $(c_\mathrm{s}^2 + c_\mathrm{A}^2 + \delta v^2 )^{0.5}$, where $c_\mathrm{s}$ is the sound speed in the medium, $c_\mathrm{A}$ is the Alfv\'{e}n speed, and $\delta v$ is the velocity dispersion. The authors proposed that the supersonic cloud collisions of at least a few km s$^{-1}$ can achieve produce a large mass accretion rate of $\sim$$10^{-4}$--10$^{-3}$ $M_\odot$ yr$^{-1}$ because of increasing the effective Jeans mass, which allows to increase stellar mass against own stellar feedback. Another numerical simulation by \citet{2014ApJ...792...63T} mentioned that diffuse gas components bridging two collided clouds in the velocity space are generated by cloud collisions, which have intermediate velocities of the two clouds. For the case of NGC~2359, the H{\sc i} component at the velocity of $\sim$41 km s$^{-1}$ corresponds to the bridging feature originated by the cloud-cloud collision which physically connects the 37 km s$^{-1}$ CO cloud to the 54 km s$^{-1}$ CO cloud.}

{A complementary spatial distribution between two colliding clouds is also expected in the sites of cloud-cloud collisions. The colliding two clouds are thought to be different morphology, mass, and column density as previously pointed out by both the observational and numerical results (e.g., \cite{1992PASJ...44..203H,2010MNRAS.405.1431A,2016ApJ...820...26F}, \yearcite{2017PASJ...69L...5F}; \cite{2018PASJ...70S..53I,2019arXiv190808404S}). For the typical case of collisions between a small and large clouds, the small cloud can create a hole in the large cloud. The collided two clouds therefore show a complementary spatial distribution if the collision angle is small respect to the line of sight. The best observational example is the Galactic super star cluster Westerlund~2 \citep{2009ApJ...696L.115F,2010ApJ...709..975O}. There are two molecular complexes toward the cluster; one is the I-shaped molecular cloud, and the other is two small clouds which are elongated in the direction perpendicular to the other cloud. The two molecular complexes are complementary distributed and form the T-shaped morphology, whose crossing point (collisional area) contains the super star cluster. For the case of NGC~2359, the 37 and 54 km s$^{-1}$ CO clouds show good complementary spatial distributions each other, whereas the WR star is located $\sim$7 pc away from each cloud. It means that the first collisional point of each cloud was different from the present cloud positions, because of the projection effect. The 37 km s$^{-1}$ CO cloud was collided with the 54 km s$^{-1}$ CO cloud toward the WR star for the first time, then the 37 km s$^{-1}$ CO cloud is continually moving to east after the cloud-cloud collision and forming the progenitor of WR star. The bent shape of the 54 km s$^{-1}$ CO cloud is also qualitatively consistent with dynamical motion of the 37 km s$^{-1}$ CO cloud, because the part of the 54 km s$^{-1}$ CO cloud is stripped or kinked by the continuous collision. The spatial separation between the North and South H{\sc i} clouds is reduced toward the eastern side, which is also possibly related to gas accumulation cased by the dynamical collision of the 37 km s$^{-1}$ CO cloud from west to east, in addition to the ionized effect due to the WR star and its progenitor. In fact, the similar shape of CO/H{\sc i} is also seen in several regions of cloud-cloud collisions (e.g., \cite{2015ApJ...806....7T}, \yearcite{2018PASJ...70S..51T}; \cite{2019ApJ...886...14F,2019ApJ...886...15T}). To summarize, the morphological and kinematical aspects of the 37 and 54 km s$^{-1}$ CO (and H{\sc i}) clouds are consistent with the observational signatures of a cloud-cloud collision.}

\subsection{Evolutionary {scheme and overall picture of the cloud-cloud collision in NGC~2359}}\label{sec:evolutionary}
\begin{figure*}
\begin{center}
\includegraphics[width=\linewidth]{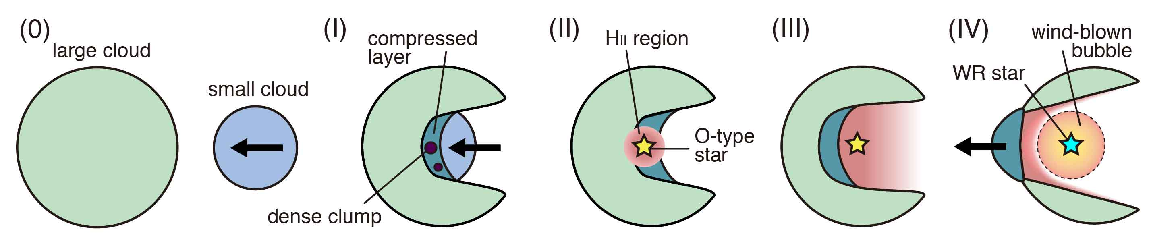}
\end{center}
\caption{Schematic illustrations of the cloud-cloud collision model presented by \citet{1992PASJ...44..203H} and by \citet{2015ApJ...806....7T}, with minor modifications. {Stages (0), (I), (II), (III), and (IV) indicate evolutional stages of the collisions: the earliest stage is (0) and the latest stage is (IV). The collisional direction of the small cloud is shown as the arrows in stages (0), (I), and (IV). For stage (0), before the collision of large and small clouds with a supersonic velocity separation. The small cloud is collided with the large cloud, and then the small cloud creates a compressed layer with dense clumps to cave in the large cloud (stage I). For stage (II), a dense clump forms an O-type star with a compact H{\sc ii} region. If the velocity of the colliding small cloud remains high, the shock-compressed layer continues to burrow into the large cloud with an evolved H{\sc ii} region (stage III) and, finally, to penetrate it (stage IV). Also, the O-type star will be evolved into the WR star with a wind-blown bubble such as NGC~2359.}}
\label{ccc_schematic}
\end{figure*}

{In this section, we first} propose {a possible scenario} that the evolutionary stage of NGC~2359 corresponds to the final phase of the cloud-cloud collision. Figure \ref{ccc_schematic} shows a schematic illustration of the cloud-cloud collision model presented by \citet{1992PASJ...44..203H} and by \citet{2015ApJ...806....7T}, with minor updates. {The authors assumed two molecular clouds with different sizes as the initial condition: one is referred to as ``large cloud'' shown in green circle, and the other referred to as ``small cloud'' shown in blue circle. The earliest stage is (0) and the latest stage of is (IV). For the first stage (0) represents the state before the collision. At the stage (I), the small cloud collids with the large cloud, and then the supersonic collision creates a compressed layer with dense clumps at the collisional front. For the stage (II), a dense clump forms an O-type star with a compact H{\sc ii} region. If the velocity of the colliding small cloud remains high, the shock-compressed layer continues to burrow into the large cloud with an evolved H{\sc ii} region (stage III) and, finally, the small cloud penetrates the large cloud (stage IV). Also, the O-type star will be evolved into the WR star with a wind-blown bubble such as NGC~2359.} {Note that} this {evolutionary scheme} is also compatible with numerical simulations of the cloud-cloud collision presented by \citet{2010MNRAS.405.1431A}. {The dense clumps in stage (I) are generated by the simulation of cloud-cloud collisions; they showed that the density is increased by at least an order of magnitude than the pre-collision cloud. The V-shaped gas distributions pointing in the direction of the collision at the stages (I--IV) as well as the distributions of colliding clouds at the stage (IV) are also reproduced by the numerical simulation (see Figure 1 in \cite{2010MNRAS.405.1431A}).}

\begin{figure*}
\begin{center}
\includegraphics[width=\linewidth]{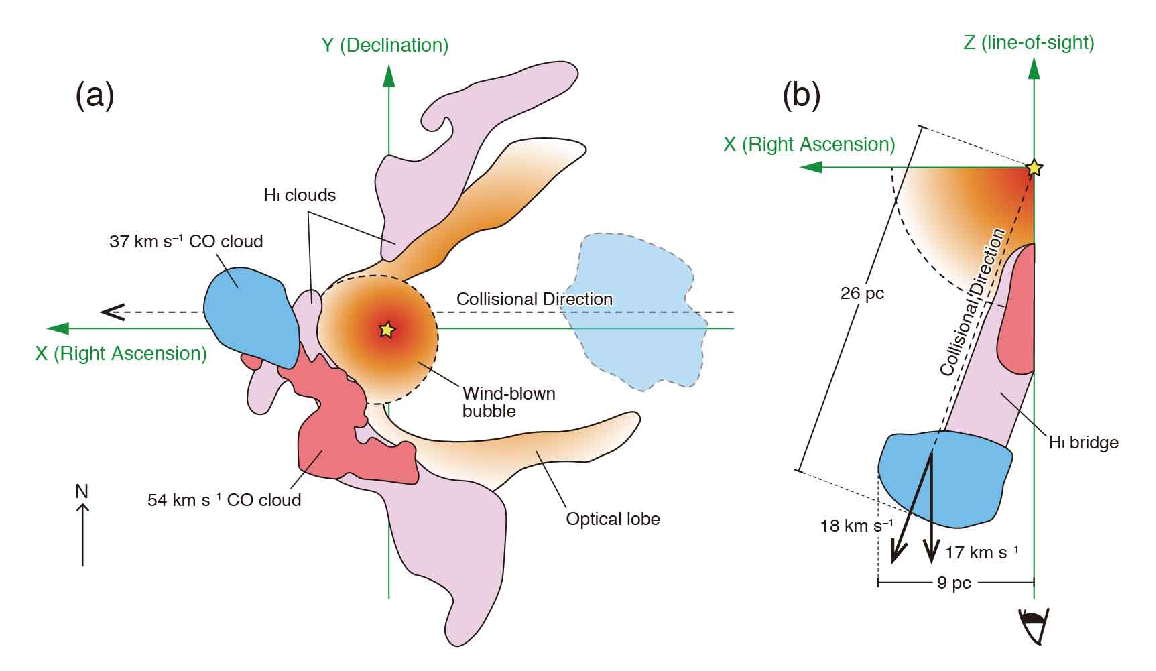}
\end{center}
\caption{Schematic illustrations of the cloud-cloud collision in NGC~2359 are presented on the {X--Y plane in (a) and X--Z plane} in (b), where the X- and Y-axes are defined as {Right Ascension and Declination}, respectively, and the Z-axes represents to along the line-of-sight. {The 37 and 54 km s$^{-1}$ CO clouds, H{\sc i} clouds (and their bridge component), position of the WR-star and its wind-blown bubble, and optical lobe(s) are also shown in the Figure.}}
\label{ngc2359_schematic}
\end{figure*}

{On the basis of the evolutionary scheme, we discuss the overall picture of cloud-cloud collision in NGC~2359.} Figure \ref{ngc2359_schematic}(a) shows a schematic illustration of NGC~2359 presented in the {Right Accession--Declination} plane. {As mentioned in Section \ref{sec:ccc}}, it is likely that the {37 km s$^{-1}$} CO cloud collided with the {54 km s$^{-1}$} CO and H{\sc i} clouds from west and that the {37 km s$^{-1}$} CO cloud {shows larger size and gas mass} than the present cloud. The collision then created an isolated O-type star in the shock-compressed layer, {and then the collided cloud} has been moved in the figure to the position of the observed the {37 km s$^{-1}$ CO cloud}. The O-type star illuminates the cavity in the 54 km s$^{-1}$ CO/H{\sc i} clouds formed by the collision, finally, transforming {the O-type star} into a WR star with a wind-blown bubble.

The collision time scale is also important for understanding the cloud-cloud collision in NGC~2359. It is thought that a WR stars is formed after passing through the main-sequence phase of an O-type star and {the red supergiant (RSG) and/or LBV phases}, the typical lifetime of which is $\sim$1.5 Myr or longer (e.g., \cite{2005A&A...429..581M}). For consistency with this time scale, the {37 km s$^{-1}$} CO cloud must have collided with at a collision angle of {20${^\circ}$ or less} relative to the {line of sight, because the collision time scale is derived as a dynamical time scale using a spatial separation between the WR star and the 37 km s$^{-1}$ CO cloud and velocity difference of two colliding clouds}. Figure \ref{ngc2359_schematic}(b) shows a schematic illustration of the cloud-cloud collision in NGC~2359 presented in the {Right Ascension--line of sight} plane. The estimated distance between the WR star and the {37 km s$^{-1}$} CO cloud is $\sim$26 pc, and the collision velocity is {$\sim$18 km s$^{-1}$}, assuming {the} collision angle of {20${^\circ}$}. {We note that} the large distance between the WR star and the {37 km s$^{-1}$} CO cloud is also consistent with the {modest} kinematic temperature {of $\sim$17 K} in the {37 km s$^{-1}$} CO cloud; it is too distant to be strongly affected by heating from the WR star. With these parameters, the {dynamical} collision timescale {can be derived to 26 pc / 18 km s$^{-1}$} $\sim$1.5 Myr. {Note that the actual collision time scale may be longer than the present estimation, because we cloud not take into account the initial sizes of the collided two clouds. Anyway, the collision time scale of CO clouds is not inconsistent with the formation and evolution time scales of the isolated O-type star, which is the progenitor of the WR star HD~56925.}

\subsection{Comparison with other {high-mass} stars formed by cloud-cloud collisions}\label{sec:comparison}
\begin{figure*}
\begin{center}
\includegraphics[width=\linewidth]{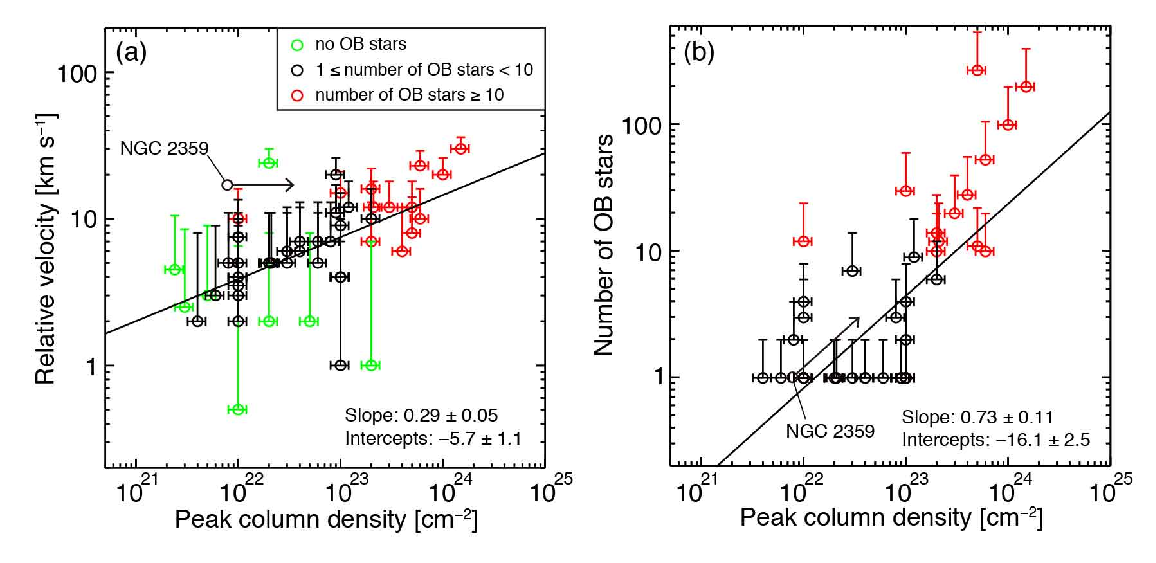}
\end{center}
\caption{{Scatter plots of (a) the peak column density of molecular hydrogen $N$(H$_2$) and relative velocity of colliding two clouds, and (b) $N$(H$_2$) and the number of OB stars toward the Galactic sources published by \citet{2019PASJ..tmp..127E}. The green, black, and red circles indicate cloud-cloud collisions associated with no OB stars, less than 10 OB stars, and more than 10 OB stars, respectively. The solid lines indicate the best-fit results using a least-squares method for each panel derived by \citet{2019PASJ..tmp..127E}. We also added the data points of NGC~2359. The allows and their directions give the lower limit of values and to be increased directions, respectively.}}
\label{ccc_summary}
\end{figure*}

{Here,} we argue that the physical properties of the colliding clouds {and the number of high-mass star in NGC~2359 are roughly consistent with that in the Galactic high-mass star(s) which were formed by cloud-cloud collisions. Figure \ref{ccc_summary}(a) shows the scatter plot between the peak column density of molecular hydrogen and relative velocity of colliding two clouds, which was summarized by \citet{2019PASJ..tmp..127E}. The authors found that the number of high-mass star(s) formed by the cloud-cloud collision will be increased with rises in both the peak column density and relative velocity difference of the colliding two clouds. For the case of NGC~2359, the relative velocity difference is slightly higher than the expected peak column density derived by the least-squares fitting for all samples. This indicates that large parts of the molecular clouds in NGC~2359 have already been ionized by strong UV radiation from the WR star, and hence the derived peak column density gives a lower limit. In other words, the initial column densities of the colliding two clouds were roughly an order of magnitude higher than the column densities of present clouds. In fact, there is a large amount of ionized gas whose total mass was estimated to be $\sim$2,000 $M_\odot$ \citep{1999AJ....118..948C}. Figure \ref{ccc_summary}(b) shows the scatter plot between the peak column density of colliding two clouds and the number of high-mass stars \citep{2019PASJ..tmp..127E}. For NGC~2359, the progenitor mass of the WR star represents two or three of high-mass stars in the samples, then we plotted it as the lower limit, as well as the peak column density. Both the plots show a nice agreement with the physical parameters of column density and relative velocity difference of colliding two clouds, and the number of high-mass star(s) formed by cloud-cloud collisions.}

{We finally present a detailed comparison with RCW~120 in order to better understand the role of the present studies of NGC~2359 for the high-mass star formation triggered by cloud-cloud collisions. The isolated O-type star in RCW~120 was also formed by collisions of two clouds, whose velocity separation is $\sim$20 km s$^{-1}$ toward the line of sight \citep{2015ApJ...806....7T}. The situation is very similar to NGC~2359, but the evolutionary stage is different. According to \citet{2015ApJ...806....7T}, the collision time scale of RCW~120 is estimated to be 0.2--0.4 Myr, which corresponds to the stage (II) or (III) in Figure \ref{ccc_schematic}. Therefore, the peak column densities of collided clouds in RCW~120 ($\sim3.2 \times 10^{22}$ cm$^{-2}$) are higher than those in NGC~2359 ($\sim0.8 \times 10^{22}$ cm$^{-2}$). In addition, the size of H{\sc ii} region in NGC~2359 is $\sim$15 pc, which is roughly five times larger than that in RCW~120. There are two possibilities to explain the difference; one is due to the different evolutionary stage as previously mentioned, and the other is difference of initial conditions of interstellar environment. For the latter case, the big difference is cased by the locations of the H{\sc ii} regions. RCW~120 is located in the inner Galaxy within 12 degrees respect to the the Galactic center, whereas NGC~2359 is placed in the near anticenter of the Galaxy where the low density H{\sc i} gas is dominated. In fact, one of the colliding clouds in RCW~120 has a large molecular mass of $5.1 \times 10^4$ $M_\odot$, which is an order magnitude higher than the total cloud mass of NGC~2359 even considering the all ionized gas mass of 2,000 $M_\odot$. The cloud density of RCW~120 derived by the LVG analysis also indicates an order magnitude higher than that of NGC~2359. Since the Str\"{o}mgren radius is inversely proportional to the gas density, the smaller size of H{\sc ii} region in RCW~120 is naturally expected. To summarize, there is no significant difference of RCW~120 and NGC~2359 in terms of high-mass star formation even if the interstellar environment is different. In other words, the strong gas compression by the cloud-cloud collision might be more effective to form an isolated high-mass star in low density environment such as outer Galaxy. To investigate the cloud-cloud collision scenario as an universal formation mechanism of high-mass star(s), we should increase the observational samples not only for the inter Galaxy, but also for the outer Galaxy. NGC~2359 makes an important evidence for the isolated high-mass star formation triggered by the cloud-cloud collision in the low density environment of the outer Galaxy.} %

\section{Conclusions} \label{sec:conc}
We have made {fully-sampled} $^{12}$CO($J$ = 1--0, 3--2) observations of {the entire} WR nebula NGC~2359 using the Nobeyama 45-m and ASTE radio telescopes. {By analyzing CO as well as archived H{\sc i} and radio continuum datasets, we proposed a possible scenario that the progenitor of} WR star HD~56925 in NGC~2359 was likely to have been formed by a cloud-cloud collision. We summarize the main results obtained in {the present} study below:

\begin{enumerate}
\item {We have confirmed presence of six molecular and atomic clouds toward the nebula: three of them are CO clouds at the velocities of $\sim$37, 54, and 67 km s$^{-1}$, and the others are H{\sc i} clouds named the North / South H{\sc i} clouds at $\sim$54 km s$^{-1}$ and the 63 km s$^{-1}$ H{\sc i} cloud. The CO/H{\sc i} clouds at 54 km s$^{-1}$ show elongated morphologies which are nicely along the southern and northern shells of radio continuum, whereas the CO cloud at 37 km s$^{-1}$ shows a clumpy distribution overlapped with the eastern radio shell. The diffuse H{\sc i} cloud at 67 km s$^{-1}$ also shows a good spatial correspondence with the shell boundary of radio continuum. The CO cloud at 67 km s$^{-1}$ has a filamentary distribution which lies in the outside of the radio shell, implying the background object. The CO/H{\sc i} clouds at 54 km s$^{-1}$ as well as the southern part of the 37 km s$^{-1}$ CO cloud are limb-brightened in radio continuum, suggesting the part of these clouds are ionized by strong UV radiation from the WR star.}

\item {The expanding CO/H{\sc i} shells are newly identified at the center velocity of $\sim$51 km s$^{-1}$, whose expanding velocity is $\sim$4.5 km s$^{-1}$. The CO cloud at 54 km s$^{-1}$ and the South H{\sc i} clouds show the large line widths and are along the expanding shell in the position--velocity diagram, suggesting both the clouds are accelerated by stellar winds from the WR star and its progenitor.}

\item {From the LVG analysis of CO clouds, we found high-kinematic temperature $T_\mathrm{kin}$ and number density of molecular hydrogen $n(\mathrm{H}_2)$ toward the 54 km s$^{-1}$ CO cloud [$T_\mathrm{kin} = 63$ K and $n(\mathrm{H}_2) = 2.6 \times 10^4$ cm$^{-3}$], suggesting that the UV heating and wind compression occurred. On the other hand, there is no heating and compression toward the 67 km s$^{-1}$ CO cloud [$T_\mathrm{kin} = 6$ K and $n(\mathrm{H}_2) = 600$ cm$^{-3}$]. The kinematic temperature of CO cloud at 37 km s$^{-1}$ ($T_\mathrm{kin} = 17$ K) is three times higher than that of the 67 km s$^{-1}$ CO cloud.}

\item {By considering the spatial distributions, kinematics, and physical properties of the CO/H{\sc i} clouds, we concluded that the 37 and 54 km s$^{-1}$ CO clouds and three H{\sc i} clouds are associated with NGC~2359, indicating that the five clouds are located at the roughly same distance from us even if the cloud velocities are different. We also found that the momentum of these clouds except for the 37 km s$^{-1}$ CO cloud can be explained by that of the stellar feedback owing to the WR star and its progenitor.}

\item {To explain the large velocity difference of the 37 km s$^{-1}$ CO cloud respect to the {systemic} velocity of NGC~2359, we proposed a possible scenario that the progenitor of WR star was formed by collisions between the 37 and 54 km s$^{-1}$ CO clouds. The three observational signatures of the cloud-cloud collision are satisfied: a supersonic velocity separation of colliding two clouds ($\sim$17 km s$^{-1}$ in NGC~2359), complementary spatial distributions of the clouds, and a bridging feature physically connecting the two clouds in the velocity space. If the scenario is correct, NGC~2359 corresponds to the final phase of the cloud-cloud collision proposed by \citet{1992PASJ...44..203H} and \citet{2015ApJ...806....7T}. The physical properties of colliding clouds and the number of formed high-mass star are roughly consistent with the previous studies in the triggered star forming regions by cloud-cloud collisions. Since NGC 2359 is located in the anticenter, the high-mass star formation triggered by cloud collisions might be more effective scheme to form an isolated high-mass star in the low-density environment such as outer Galaxy.}
\end{enumerate}

\begin{ack}
The authors would like to thank Robert Franke from the Focal Pointe Observatory for providing the optical images of NGC~2359. The NANTEN project is based on a mutual agreement between Nagoya University and the Carnegie Institution of Washington (CIW). We appreciate the hospitality of all the staff members of the Las Campanas Observatory of CIW. We are thankful to many Japanese companies and public donors who contributed to the realization of the project. The ASTE telescope is operated by  National Astronomical Observatory of Japan (NAOJ). The Nobeyama 45-m radio telescope is operated by Nobeyama Radio Observatory, a branch of NAOJ. {The FUGIN data were retrieved from the JVO portal (http://jvo.nao.ac.jp/portal/) operated by ADC/NAOJ.} This work was financially supported by Grants-in-Aid for Scientific Research (KAKENHI) of the Japanese society for the Promotion of Science (JSPS, grant No. 15H05694). This work also was supported by ``Building of Consortia for the Development of Human Resources in Science and Technology'' of Ministry of Education, Culture, Sports, Science and Technology (MEXT, grant No. 01-M1-0305). We acknowledge to Kosuke Fujii and Hiroaki Iwamura for contributions on the observations of $^{12}$CO($J$ = 3--2) and $^{12}$CO($J$ = 1--0) data. 
\end{ack}

\section*{{Appendix: Determination of CO-to-H$_2$ conversion factor}} \label{sec:appendix}
{To derive the CO-to-H$_2$ conversion factor $X$(CO) toward the NGC~2359 region, we used the NANTEN $^{12}$CO($J$ = 1--0), {\it{Planck}} {dust optical depth} $\tau_\mathrm{353}$ at the frequency of 353~GHz, and the {\it{Planck}} dust temperature $T_\mathrm{d}$ maps following the method published by \citet{2017ApJ...838..132O}.}

\begin{figure*}
\begin{center}
\includegraphics[width=\linewidth]{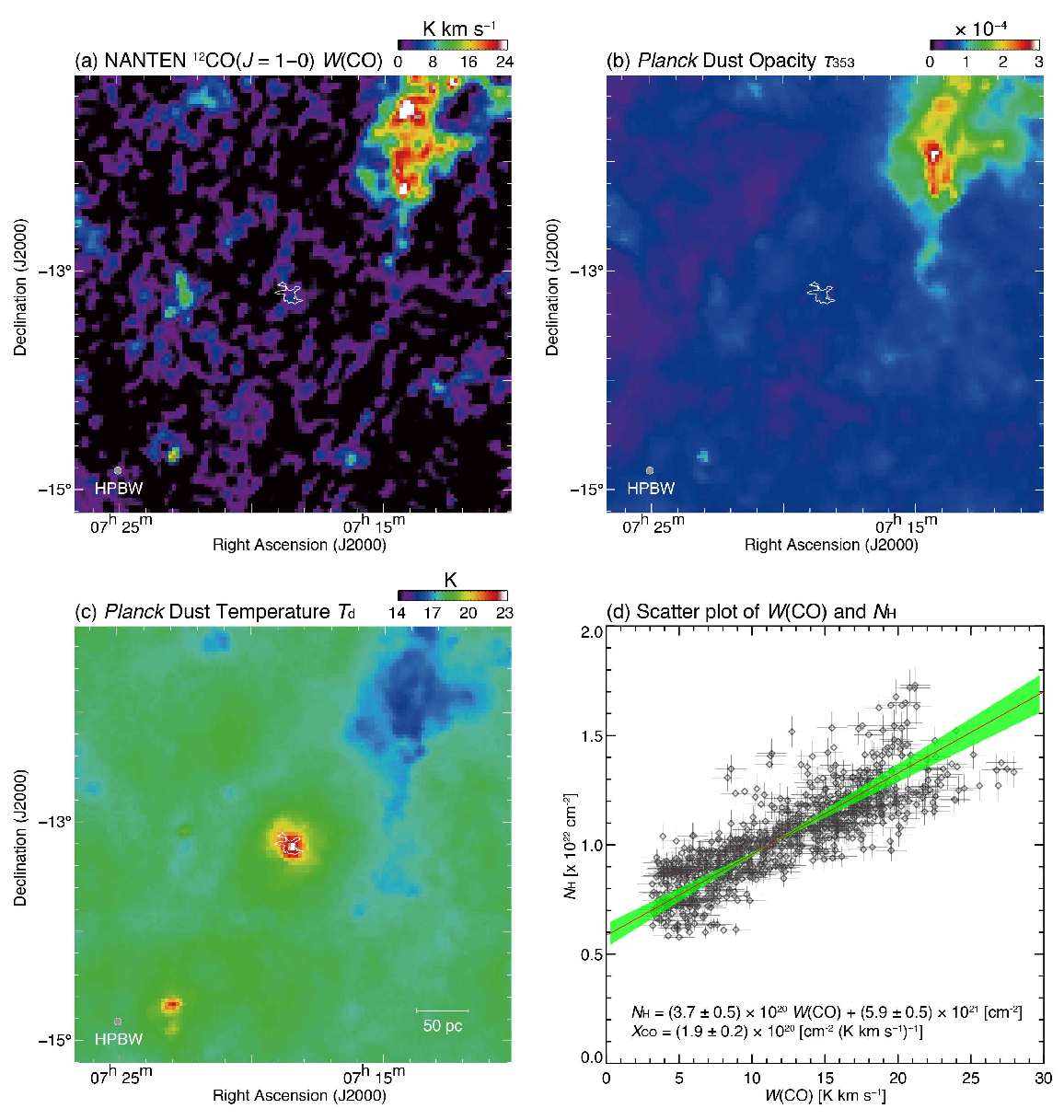}
\end{center}
\caption{{Total intensity maps of (a) NANTEN $^{12}$CO($J$ = 1--0), (b) {\it{Planck}} {dust optical depth} $\tau_\mathrm{353}$ at the frequency of 353~GHz, and (c) {\it{Planck}} dust temperature $T_\mathrm{d}$. The integration velocity range of CO is from 5 to 45 km s$^{-1}$, which contains all CO emission toward the region.} Size of each map is 5 degree $\times$ 5 degree. {The superposed contours indicate the VLA radio continuum boundary as shown in Figure \ref{channelmap}}. The beam size of each map is also shown in the left bottom corner. (d) {Scatter} plot between the total integrated CO intensity $W$(CO) and the total proton column density $N_\mathrm{H}$. {The one-sigma error bars are also plotted. The red line indicates the linear regression of the plot applying a least squares fit using the MPFITEXY procedure of IDL. The area colored in green shows variation of the correlation provided by the least squares fit. The derived equation and $X$(CO) factor are also shown in the bottom panel.}}
\label{xfactor}
\end{figure*}

{Figures \ref{xfactor}(a--c) show the total intensity maps of CO, $\tau_\mathrm{353}$, and $T_\mathrm{d}$ covering 5$^{\circ}$ $\times$ 5$^{\circ}$ around the WR nebula NGC~2359.} The nebula, indicated by the white contours of the radio shell boundary, is located in the direction of a low-density region. A prominent giant molecular cloud is located at {($\alpha_\mathrm{J2000}$, $\delta_\mathrm{J2000}$) $\sim$ ($07^{\mathrm{h}}14^{\mathrm{m}}$, $-12{^\circ}00\arcmin$), but is faraway from NGC~2359 with a spatial separation of $\sim$200 pc assuming the distance of 5 kpc}. {We note that} the {regions} of CO {$> 8$ K km s$^{-1}$}, $\tau_\mathrm{353} > 1 \times 10^{-4}$, and $T_\mathrm{d} < 17$ K {show} a good spatial correspondence {each other}. Similar trends have been found in previous studies of the high-latitude cloud complexes {(MBM~53, 54, 55, and HLCG~92$-$35, \cite{2014ApJ...780...36F,2018PASJ..tmp..132F}; Perseus, \cite{2017ApJ...838..132O}; Chamaeleon, \cite{2019ApJ...878..131H,2019ApJ...884..130H}) and the near Galactic supernova remnant of RCW~86 \citep{2019ApJ...876...37S}.}

According to \citet{2017ApJ...838..132O}, the total interstellar hydrogen column density $N_\mathrm{H}$ is given by;
\begin{eqnarray}
N_\mathrm{H} = 9.0 \times 10^{24} \cdot (\tau_\mathrm{353})^{1/1.3},%\phantom{0}(\mathrm{cm}^{-2}),
\label{eq1}
\end{eqnarray}
where the non linear exponent 1/1.3 includes the effect of the dust-growth parameter discussed by \citet{2013ApJ...763...55R} and \citet{2017ApJ...838..132O}. {Here, $N_\mathrm{H}$ also can be written by the following equations,}
\begin{eqnarray}
N_\mathrm{H} = 2 X(\mathrm{CO}) \cdot W(\mathrm{CO}) + N(\mathrm{H}{\textsc{i}}),\\
\label{eq_xco11}
\equiv  (\mathrm{slope}) \cdot W(\mathrm{CO}) + (\mathrm{intercept}),
\label{eq_xco1}
\end{eqnarray}
{where $W(\mathrm{CO})$ is the total integrated intensity of CO and $N${(H{\sc i})} is the proton column density of atomic hydrogen. Therefore, we can derive $X(\mathrm{CO})$ by a liner fitting of the scatter plots between $N_\mathrm{H}$ and $W(\mathrm{CO})$.}

Figure \ref{xfactor}(d) shows the correlation plot between $W$(CO) and $N_\mathrm{H}$ derived using Equation (\ref{eq1}). {We perform a linear fitting using the MPFITEXY procedure of IDL}, which provides the slope and intercept by using a $\chi^2$ test \citep{2010MNRAS.409.1330W}. {We note that using only the data points with $W$(CO)$> 3 \sigma$ and $T_\mathrm{d} < 17$ K give a best fit parameters with a reduced $\chi^2 = 1.006$ at the degree of freedom of 889. We obtained the slope of (3.7 $\pm$ 0.5) $\times$ 10$^{20}$ and the intercept of (5.9 $\pm$ 0.5) $\times$ 10$^{21}$. Finally, we derived  $X(\mathrm{CO}) = 1.9 \times$ 10$^{20}$ cm$^{-2}$ (K km s$^{-1}$)$^{-1}$ using the Equation (\ref{eq_xco1}).}

\newpage

\begin{table*}
\tbl{Physical properties of CO and H{\sc i} clouds toward NGC~2359}{%
\begin{tabular}{lccccccccc}
\hline\noalign{\vskip3pt} 
\multicolumn{1}{c}{Cloud Name} & $\alpha_{\mathrm{J2000}}$ & $\delta_{\mathrm{J2000}}$ & $T_R^*$ & $V_\mathrm{peak}$  & $\Delta V_\mathrm{LSR}$ & Size & Column Density & Mass & Comment\\
 & ($^{\mathrm{h}}$ $^{\mathrm{m}}$ $^{\mathrm{s}}$) & ($^{\circ}$ $\arcmin$ $\arcsec$) & (K) & (km s$^{-1}$) & (km s$^{-1}$) & (pc) & ($\times 10^{21}$ cm$^{-2}$) & ($M_\mathrm{\odot}$) & \\  [2pt] 
\multicolumn{1}{c}{(1)} & (2) & (3) & (4) & (5) & (6) & (7) & (8) & (9) & (10) \\  [2pt] 
\hline
\hline\noalign{\vskip3pt} 
37 km s$^{-1}$ CO cloud &  07 18 44.8 &  $-$13 12 30 & \phantom{0}8.0 & 37.0 & 1.5 & \phantom{0}5.8 & 3.1 & 630 & Peak~A\\
54 km s$^{-1}$ CO cloud &  07 18 36.6 &  $-$13 17 00 & \phantom{0}9.4 & 53.4 & 4.1 & \phantom{0}5.8 & 7.8 & 890 & Peak~B\\
67 km s$^{-1}$ CO cloud &  07 18 44.8 &  $-$13 18 30 & \phantom{0}3.8 & 66.3 & 1.3 &  \phantom{0}5.0 & 1.1 & 240 & Peak~C\\
\hline\noalign{\vskip3pt} 
North H{\sc i} cloud & 07 16 05.1 & $-$13 03 20 & 15.2  & 53.2 & 3.5 & \phantom{0}8.6 & 0.4 & 130 & \\
South H{\sc i} cloud & 07 16 11.9 & $-$13 11 40 & 23.9 & 55.4 & 6.1 & 17.9 & 0.8 & 660 & \\
63 km s$^{-1}$ H{\sc i} cloud & 07 16 22.9 & $-$13 12 50 & 17.5 & 63.7 & 3.3 & 23.9 & 0.4 & 990 & \\
\hline
\hline\noalign{\vskip3pt} 
\end{tabular}}\label{table:extramath}
\begin{tabnote}
Note. --- Col. (1): Cloud name. Cols. (2--3) Equatorial coordinates of the maximum intensity of $^{12}$CO($J$ = 1--0) or H{\sc i} for each component. Cols. (4--6) Physical properties of the $^{12}$CO($J$ = 1--0) and H{\sc i} obtained at each position. Col. (4): Peak radiation temperature, $T_R^{\ast} $. Col. (5): $V_{\mathrm{peak}}$ derived from a Gaussian fitting. Col. (6): line width of full-width half-maximum (FWHM), $\bigtriangleup V_{\mathrm{LSR}}$. Col. (7): CO/H{\sc i} cloud size defined as ($A$/$\pi$)$^{0.5} \times 2$, where $A$ is the total cloud surface area surrounded by the 3 $\sigma$ level in the integrated intensities of the CO/H{\sc i} cloud. Col. (8): For the CO clouds, the molecular hydrogen column density $N$(H$_2$) derived from the $^{12}$CO($J$ = 1--0) integrated intensity, $W$(CO), and $N$($\mathrm{H_2}$) = 1.9 $\times $  $10^{20}$ [$W$(CO)/(K km $\mathrm{s^{-1}}$)] ($\mathrm{cm^{-2}}$) (see in the text). For the H{\sc i} clouds, the atomic hydrogen column density $N$(H{\sc i}) derived using the equations of {$N$(H{\sc i})} = 1.823 $\times$ 10$^{18}$ $W$(H{\sc i}). Col. (9): Mass of the cloud derived using the relation between the molecular or atomic hydrogen column density, shown in Col. (8). Col. (10): The names of peaks A--C used for the LVG analysis (see the text).
\end{tabnote}
\end{table*}

\end{document}